\documentclass[twocolumn]{aastex701}
\usepackage{graphicx} 
\usepackage{amsmath}
\usepackage{natbib}
\usepackage{multirow}
\usepackage{makecell}
\usepackage{verbatim}

\usepackage{xcolor}
\definecolor{new_color}{HTML}{CF0000}
\definecolor{new_green}{HTML}{e69b1c}

\begin{document}

\title{Planets surviving common envelope evolution should be rare, massive, and orbit very close to their host star}

\correspondingauthor{Yadira S. Gaibor} 
\email{ygaibor@mit.edu}

\author[0000-0002-2036-2311]{Yadira S. Gaibor}
\email{ygaibor@mit.edu}
\affiliation{Department of Physics and Kavli Institute for Astrophysics and Space Research, Massachusetts Institute of Technology, Cambridge, MA 02139}

\author[0000-0001-7246-5438]{Andrew Vanderburg}
\email{avanderburg@cfa.harvard.edu}
\affiliation{Department of Physics and Kavli Institute for Astrophysics and Space Research, Massachusetts Institute of Technology, Cambridge, MA 02139}
\affiliation{Center for Astrophysics Harvard \& Smithsonian, 60 Garden Street, Cambridge, MA 02138, USA}

\author[0000-0002-7733-4522]{Juliette C. Becker}
\email{juliette.becker@wisc.edu}
\affiliation{Department of Astronomy,  University of Wisconsin--Madison, 475 N.~Charter St., Madison, WI 53706, USA}

\author[0000-0001-7493-7419]{Melinda Soares-Furtado}
\email{mmsoares@wisc.edu}
\affiliation{Department of Astronomy,  University of Wisconsin--Madison, 475 N.~Charter St., Madison, WI 53706, USA}
\affiliation{Department of Physics, University of Wisconsin--Madison, 1150 University Ave., Madison, WI 53706, USA}

\author[0000-0003-3182-5569]{Saul Rappaport}
\email{sar@mit.edu}
\affiliation{Department of Physics and Kavli Institute for Astrophysics and Space Research, Massachusetts Institute of Technology, Cambridge, MA 02139}

\author[0000-0002-6939-9211]{Tansu Daylan}
\email{tansu@wustl.edu}
\affiliation{Department of Physics and McDonnell Center for the Space Sciences, Washington University, St. Louis, MO 63130, USA}

\begin{abstract}

The discovery of exoplanets around white dwarfs uniquely informs our understanding of planetary survival through stellar evolution. Common Envelope (CE) evolution is one of the proposed mechanisms that could shape the orbital architecture of these planets around stellar remnants. 
In this work, we predict the occurrence rate and orbital separation/planet mass distributions for planets that could form around white dwarfs through CE evolution.
We simulate populations of giant-planet systems, sampling initial conditions from observed mass and semi-major axis distributions for main sequence stars. We evolve the host star from main sequence to white dwarf to assess which planetary configurations could survive, and we predict the resulting post-CE planet population. 
Our results suggest that such planets should be rare, massive, and have very close orbital separations. We find that most of the surviving planets have masses of $>10\,M_{\rm{Jup}}$ and orbital separations $<1.4$ times the tidal disruption radius, $a_{{disrupt}}$ ($\lesssim0.006$\,AU, also called the  Roche limit), which is in contrast to high eccentricity migration, which produces planets with orbital separations $>2a_{disrupt}$. The properties of CE survivors from our simulations contrast with the known closely-orbiting planet WD 1856+534 b, suggesting a different mechanism may be responsible for its formation.
\end{abstract}

\section{Introduction}

The fate of exoplanets is tied to the evolution of their environments, including the life cycle of their host stars. It has been established that exoplanets are present at different stages of the evolution of the host star, including its death. 
Observational evidence for planetary survival in the late stages of stellar evolution includes planets around pulsars \citep[e.g.][]{Wolszczan92}, accreting planetary material onto polluted white dwarfs \citep[e.g.][]{Zuckerman10, Koester14, Hollands17}, and some intact planet candidates around white dwarfs \citep[e.g.][]{Luhman11, Gansicke19, Zhang24, Blackman21}.
Three of these intact candidates are in close orbits around their star (WD 1856+534 b, \citealt{Vanderburg20}; WD 0310-688 b, \citealt{limb24}; and WD J0914+1914 \citealt{Gansicke19}). WD 1856b was recently confirmed to be a planet with a mass between $1-6\,M_{\rm{Jup}}$ \citep{limbach25} and an orbital period of 1.4 days \citep{Vanderburg20}.

The main challenge of explaining the presence of planets in close orbits around stellar remnants is that the planet has to survive the final stages of the star's post-main-sequence evolution, specifically the Red Giant branch (RGB) and Asymptotic Giant Branch (AGB). Planets with larger orbital separations should be able to avoid engulfment and are expected to survive the giant stages of their host star. In contrast, the present-day orbits of close-in planets around white dwarfs, like WD 1856b, lie well within the maximum radius reached by the host star during its giant phases. This implies that either the planets survived engulfment or migrated inward after the white dwarf formed.

That challenge has made WD 1856b the subject of considerable speculation regarding its formation history \citep{OConnor20, Stephan21, Munoz20, Lagos21, chamandy21, merlov21}. Two primary formation pathways have been proposed: common envelope (CE) evolution and high-eccentricity (HE) migration. HE migration involves an initial excitation of the planet's orbit to a high eccentricity, potentially via Kozai-Lidov oscillations, planet-planet scattering, or stellar flybys, followed by tidal circularization to a close orbit.
Although high eccentricity migration post-RGB or -AGB phase could, in principle, produce the observed configuration of systems hosting intact giant planets such as WD 1856b (as well as perturb smaller bodies, e.g. \citealt{Li2025}), there is still not enough observational evidence to tell whether this mechanism is an active formation channel \citep{OConnor20}.

In the CE scenario, the planet initially orbits a progenitor star that undergoes expansion during its red giant phase, leading to engulfment and subsequent orbital decay due to dynamical friction within the envelope. Once orbiting within the giant star's envelope, the planet must deposit enough energy into the gas to eject the star's outer layers and end migration before the planet is destroyed.
Since early works and observations on CE evolution \citep[and others that followed]{Paczynski76, Bond76}, this mechanism has become widely accepted as the explanation for the presence of close-in stellar companions to white dwarfs. However, it is less clear whether the process can extend to planetary companions. Because planets have much lower masses than stars, they release less orbital energy during inspiral, making it harder to unbind and eject the envelope. However, previous studies have concluded that CE evolution for planetary-mass companions is borderline energetically feasible \citep{Lagos21,Yarza2023}, especially if additional energy sources, like recombination energy and orbital energy from other planetary companions can contribute \citep{chamandy21}. This could make it possible for CE to form close-in planets around white dwarfs.

While a number of studies have investigated the orbital evolution and consequent survivability of planets during CE evolution, these works have focused on individual systems or on the mechanisms of survival. In this paper, we focus on predicting the observable population of planets that could emerge if CE evolution is responsible for producing close-in planets around white dwarfs. 
Other studies \citep{nord13, mauch26} have performed similar analyses, but they mostly focused on planets that survived beyond the outer reaches of the envelope. Here, we focus on the planets that get engulfed, see which of them can survive, and predict the properties of the post CE population.

In this work, we estimate the number of planets that could result from CE evolution and predict the post-CE planet population distributions of semi-major axis and planet mass. We simulate a population of giant-planet hosts drawn from observed exoplanet mass and orbital semi-major axis distributions. We vary certain simulation parameters, like the tidal $Q$ value in our tidal prescription, the efficiency of envelope ejection $\alpha$ in the CE prescription, and the host star mass.
We find that planets surviving CE evolution should be rare, massive, and orbit near their tidal disruption radii (the Roche limit). Although WD 1856b is the only confirmed example of a closely-orbiting giant planet around a white dwarf, its rarity may be reflect the incompleteness of WD surveys rather than true scarcity. Therefore, the characteristics of our resulting CE population stand in contrast to those of WD 1856b, suggesting it may have formed by a different method.

This paper is organized as follows. In Section~\ref{methods}, we describe our methods for simulating the population of giant planet hosts. We also discuss the tidal evolution and CE prescription that we use to evolve the systems, as well as the termination conditions for our simulations. Section~\ref{results} details the trends in the surviving populations for each simulation case study. In Section~\ref{discuss}, we discuss the occurrence rates of the surviving planets and compare them to the known and candidate white dwarf planets to determine how well their observed properties match the predictions of the CE scenario.
Section \ref{conclude} summarizes our main findings.

\section{Methods} \label{methods}
\begin{figure*}[!htb]
    \centering
	\includegraphics[width=0.95\textwidth]{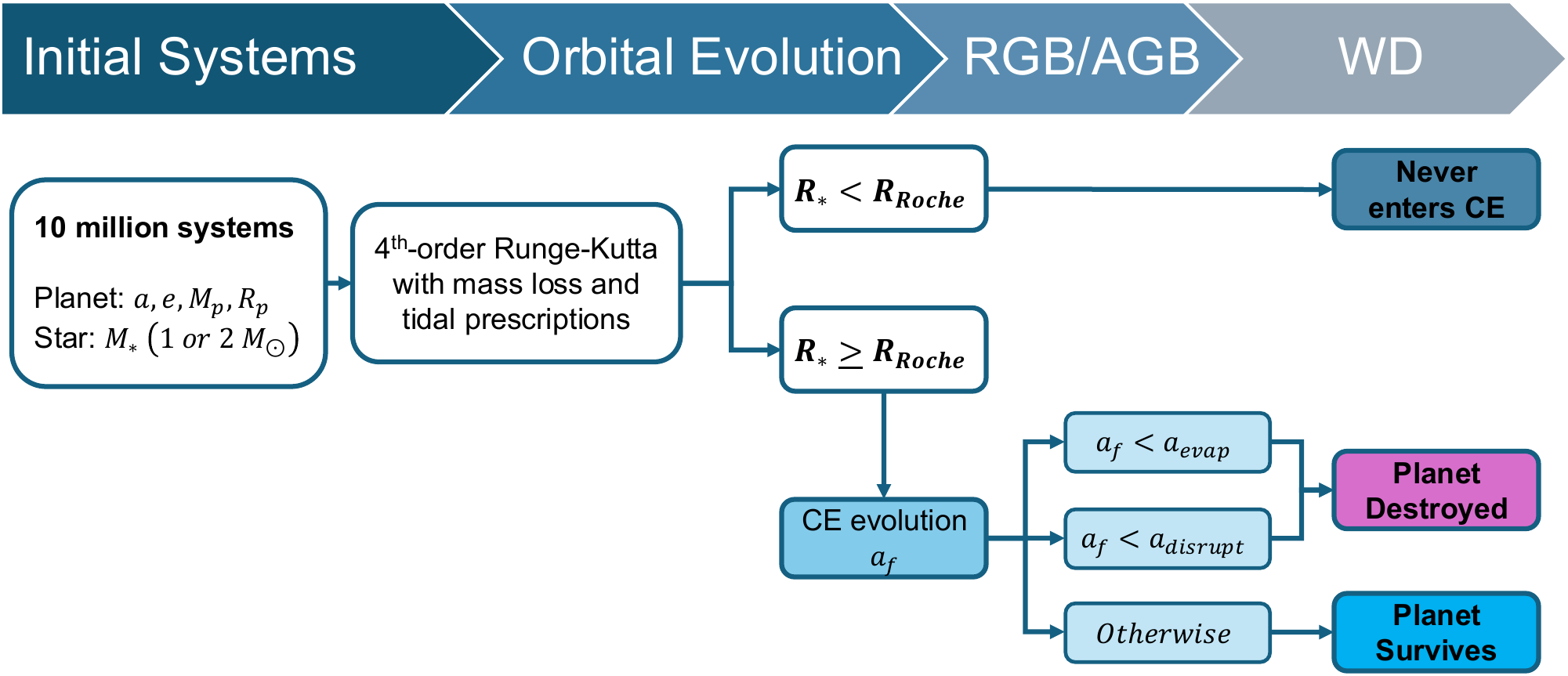}
	\caption{Summary of the steps in our simulation process. We include an approximate timeline of when the conditions for the common envelope can occur along the host star's evolution. The main parameters we compare to determine if a planet undergoes CE are the stellar radius ($R_{\star}$) and the star's Roche radius ($R_{Roche}$). We present the termination and survival conditions for planets that reach the final semi-major axis ($a_f$) after CE. 
	\label{sim_steps}}
\end{figure*}

To predict the population-level characteristics of planets that migrated through CE evolution, we ran simulations that evolve single-planet systems along with their host stars. Our approach consists of generating a population of planets drawn from observationally motivated distributions and coupling their orbital evolution to the host star's evolution from the main-sequence to the white-dwarf stage. We conduct several simulation ``runs,'' each with a different combination of assumptions to test the robustness of our predictions. We visually summarize the steps of our full evolutionary sequence in Figure~\ref{sim_steps} and provide an in-depth description of each element of our model in the following subsections.

\subsection{Initial System Parameters}\label{sec:initparams}

\begin{figure*}[!htb]
    \centering
	\includegraphics[width=0.85\textwidth]{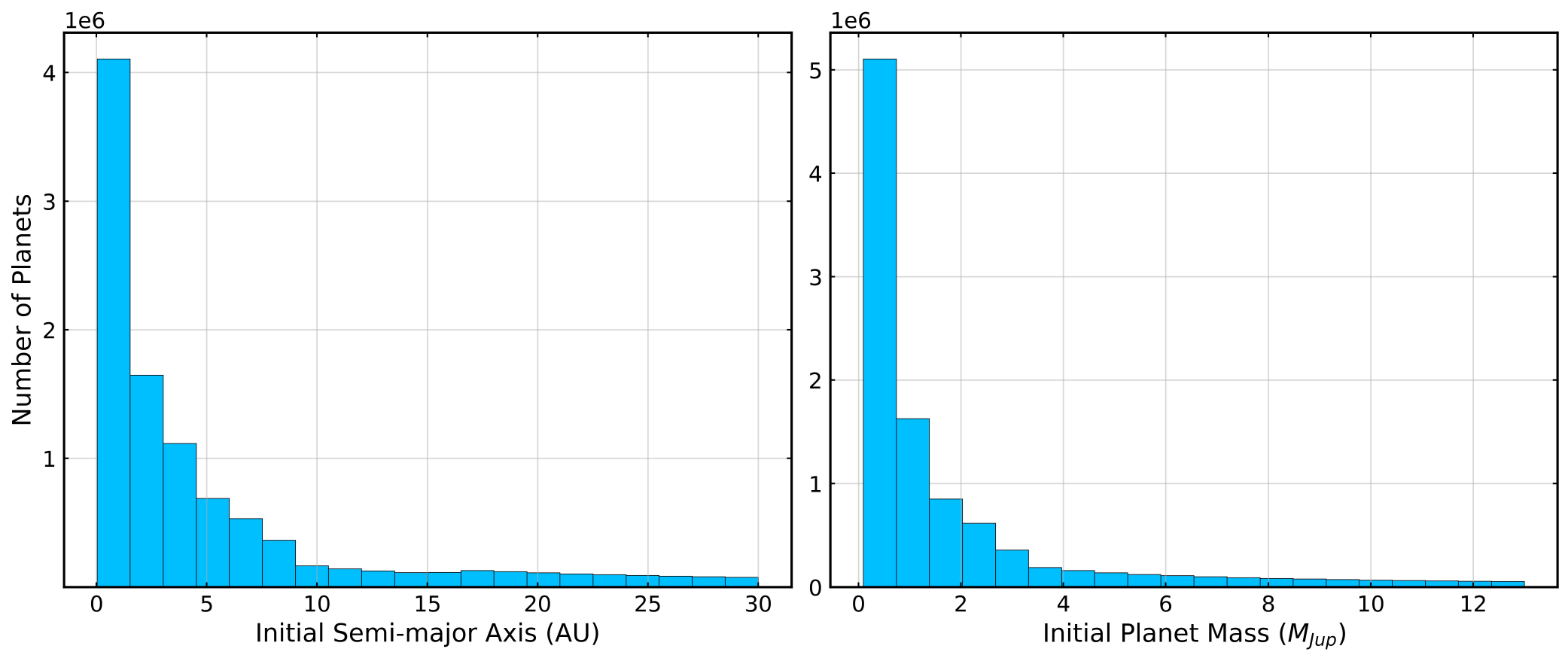}
	\caption{Initial distributions of 10 million randomly drawn planets. We ran our simulations only for select systems out of these random draws, those with $M_{\rm{p}}>2\,M_{\rm{Jup}}$ and $1\,\mathrm{AU} < a < 15\,\mathrm{AU}$ to optimize computation time.
    \label{init_distribs}}
\end{figure*}

For each run, we randomly generated 10 million planets and evolved the systems with different tidal prescriptions, stellar masses, or CE efficiency parameters. The planet's initial parameters were randomly drawn from known giant planet distributions. The mass $M_p$ and semi-major axis $a$ were sampled from the California Legacy Survey (CLS) distributions \citep{Fulton2021}. 
We sampled the initial masses from $0.1$ to $13\,M_{\rm{Jup}}$ and semi-major axis from $0.03$ to $30$\,AU, but chose to optimize our computation time by only running our full evolutionary sequence for planets likely to survive CE.
In particular, we only ran simulations for planets with mass $>2\,M_{\rm{Jup}}$ and $a$ between 1 and 15\,AU. Preliminary tests confirmed that there were no CE survivors outside of these bounds for most of our scenarios. 
These initial tests also showed that planets below $2\,M_{\rm{Jup}}$ and 1 AU never survived; therefore, we immediately mark these planets as destroyed.
For planets with $a>$ 15\,AU, initial tests showed that they never go through CE; therefore, we mark such planets as non-CE survivors.
We note that for the cases with $\alpha=$20, we relaxed the inner cutoff on $a$ and evolved planets with $a<$ 15\,AU, because initial tests showed survivors down to smaller separations.
The full initial planet mass and semi-major axis distribution are shown in Figure~\ref{init_distribs}. 
The eccentricity values $e$ were drawn from the distribution found by \citet{Kipping2013}. We calculated the planet radius $R_p$ from the \citet{ChenKipping2016} mass-radius relationship. Each planet was assigned to a host star with a main-sequence mass chosen to be either 1 or 2\,$M_{\odot}$. We chose 2\,$M_{\odot}$ for our realistic and illustrative cases because stars of this mass are the most common progenitors of white dwarfs.

\subsection{MESA Models}
We computed stellar evolution models for initial masses of 1.0 and 2.0\,$M_{\odot}$ at Solar metallicity ($Z=0.02$) using Modules for Experiments in Stellar Astrophysics (\texttt{MESA}, version r24.08.1, \citealt{pax11, pax13, pax15, pax18, pax19, jermyn23}).
Each model was evolved from the pre-main-sequence through the thermally pulsating asymptotic giant branch (TP-AGB) and subsequent white dwarf cooling.
We adopted the GS98 solar composition \citep{1998SSRv...85..161G} with Type 2 OPAL opacities.
The energy equation was solved with the \texttt{dedt} option during core hydrogen burning, switching to the \texttt{eps\_grav} formulation for all subsequent phases to improve numerical stability through the helium core flash and AGB thermal pulses. 
The \texttt{gold2} solver tolerances were enabled throughout to maintain tight energy conservation.
 
Mass loss was treated with the \citet{Reimers1975} prescription on the red giant branch and the \citet{Bloecker1995} prescription on the asymptotic giant branch, with the transition governed by a central hydrogen mass fraction threshold of $10^{-4}$. 
Angular momentum transport was included via rotationally induced instabilities, with the Tayler--Spruit dynamo, Eddington--Sweet circulation, Goldreich--Schubert--Fricke instability, and secular shear all enabled. 
During the AGB phase, we applied exponential convective overshoot \citep{Herwig2000} at all convective boundaries with $f_\mathrm{ov} = 0.014$ and $f_0 = 0.004$, and employed additional mesh refinement in and around overshoot regions to resolve thermal pulses. The AGB phase was terminated when the envelope mass fell below $10^{-2}\,M_\odot$.

\subsection{Orbital Evolution}\label{sec_orbevol}

Once we initiated the population, we evolved the planet's semi-major axis $a$ and eccentricity $e$ with a 4th-order Runge-Kutta integration method. We oversampled the \texttt{MESA} time array by a factor of 4 to obtain shorter timesteps and improve the accuracy of the time-evolution calculations. The change in semi-major axis, $da/dt$, had two components: one due to tides and one due to stellar mass loss. The eccentricity only changed due to tides:
\begin{equation}
    \begin{aligned}
    \frac{da}{dt} &= \left.\frac{da}{dt} \right|_{\mathrm{tides}} + \left. \frac{da}{dt} \right|_{\mathrm{mass}}\\
    \frac{de}{dt} &= \left. \frac{de}{dt} \right|_{\mathrm{tides}}
    \end{aligned}
\end{equation}
The change in $a$ due to stellar wind mass loss is given by:
\begin{equation}
    \left. \frac{da}{dt} \right|_{\mathrm{mass}} = -a \frac{\dot{M}_{\star}}{M_{\star}}
\end{equation}
where $M_{\star}$ is the stellar mass at the corresponding timestep and $\dot{M_{\star}}$ is the stellar mass loss rate computed directly by \texttt{MESA} rather than calculated manually, which minimizes numerical errors.
We also expected that tides raised on both the planet's and star's surface would play an important role in the planet's orbital evolution. To model the tidal evolution of orbits, we used the equilibrium tide formalism of \citet{Goldreich} to second order in eccentricity, written as in \citet{nordhaus10}:
\begin{equation}
    \begin{aligned}
    \left. \frac{da}{dt}\right|_{\mathrm{tides}} & = -\frac{1}{a^{11/2}} \left[2 \frac{K_1 R_p^5 e^2}{Q'_p} +\left( \frac{8}{25}+\frac{456e^2}{100}\right) \frac{K_2 R_{\star}^5}{Q'_{\star}}\right]\\
    \left. \frac{de}{dt} \right|_{\mathrm{tides}} & = -\frac{e}{a^{13/2}} \left[K_1 \frac{R_p^5}{Q'_p} + K_2 \frac{R_{\star}^5}{Q'_{\star}}\right]
    \end{aligned}
\end{equation}
where $R_p$ and $R_{\star}$ are the radii of the planet and the star at a given time, $Q'_p$ and $Q'_{\star}$ are the modified tidal dissipation factors of the planet and star, respectively. This low-eccentricity approximation describes well the evolution expected for the majority of our sample. For the small portion of our sample that are initialized with larger eccentricities, we may be slightly underestimating tidal dissipation, but the computed orbital evolution will be qualitatively accurate across all tested values of eccentricity. 
The constants $K_1$ and $K_2$ are given by:
\begin{equation}
    \begin{aligned}
    K_1 &= \frac{63}{4}(GM_{\star})^{1/2} \frac{M_{\star}}{M_\mathrm{p}},\\
    K_2 &= \frac{225}{16}(GM_{\star})^{1/2} \frac{M_\mathrm{p}}{M_{\star}}
    \end{aligned}
\end{equation}

\begin{figure}[!ht]
    \centering
	\includegraphics[width=\columnwidth]{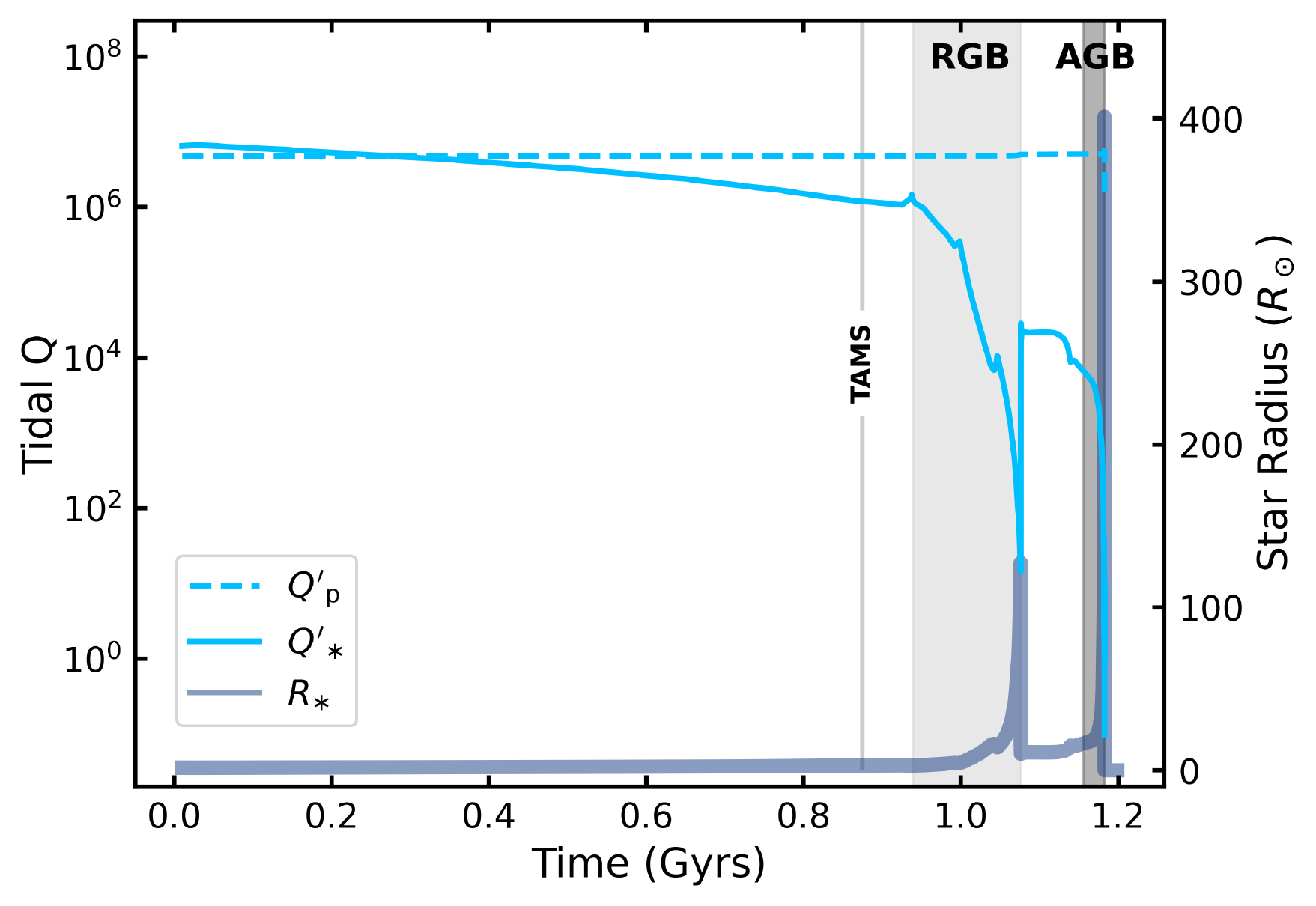}
	\caption{Evolution of variable tidal Q factors for a $2M_{\odot}$ host star. The light blue lines show the evolution of $Q'_{\star}$ (solid) and $Q'_p$ (dotted) for a planet that survives CE evolution. Critical stages of stellar evolution are labeled and shaded in gray.
	\label{tidalQevol}}
\end{figure}

We ran simulations with both a constant and a variable star and planet tidal quality factor $Q'_\mathrm{{\star,p}}$. These modified tidal factors are given by $3Q/2k_2$, where $Q$ is the specific tidal dissipation function, and $k_2$ is the Love number. The stellar static tidal parameter $Q'_\mathrm{{\star}}$ was taken to be $10^5$ or $10^6$, which are within the range of accepted values in the literature and are based on observational constraints \citep{ogil07,jack08}. The planet static tidal parameter $Q'_{\mathrm{p}}$ was taken to be $10^{5.5}$. The variable $Q'$s were calculated at each timestep, where the subscripts $\star,\rm{p}$ stand for star or planet following \citet{FabTre2007, Hansen2010}:
\begin{equation}
    \begin{aligned}
    Q'_{\mathrm{\star,p}} &=\frac{4}{3}\frac{k_{\mathrm{\star,p}}}{(1+2k_{\mathrm{\star,p}})^2}\frac{G M_{\mathrm{\star,p}}}{R_{\mathrm{\star,p}}^3}\frac{t_{V_{\mathrm{\star,p}}}}{n},\\
    n &= \sqrt{\frac{G (M_{\star} + M_\mathrm{p})}{a^3}}
    \end{aligned}
\end{equation}
where the apsidal motion constant $k_{\mathrm{\star,p}} = 0.5 \frac{Q_{\mathrm{E\star,p}}}{1-Q_{\mathrm{E\star,p}}}$ is related to the \citet{eggle98} coefficient $Q_\mathrm{E}$, for which we used 0.028 for stars and 0.30 for gas giants. The viscous timescale $t_{V_{\mathrm{\star,p}}}$ was taken to be 1.5 years for stars and 0.01 years for planets \citep{naoz12, petro15}. Figure~\ref{tidalQevol} shows the time evolution of variable $Q'_{\star}$ and $Q'_\mathrm{p}$ in one of our simulations as the planet and star evolve.

For the stages of evolution where the host star has a significant convective envelope, the viscous time scale had to be modified when the eddy turnover timescale  \citep{rasio96} became greater than half the orbital period of the planet (i.e., when $\tau_\mathrm{d}>P/2$) \citep{zahn89,villaver09}. We calculated this effective viscous timescale $t_{V_\mathrm{eff}} = t_{V}/f $, where: 
\begin{equation}
    \begin{aligned}
    \tau_\mathrm{d} = \left[ \frac{M_\mathrm{env}(R_{\star}-R_\mathrm{env})^2}{3L_{\star}} \right]^{1/3} ,\quad
    f= \left ( \frac{P}{2 \tau_\mathrm{d}} \right)^2
    \end{aligned}
\end{equation}
$R_{env}$ is the radius at the bottom of the convective envelope, and we defined the core boundary within the radial profile of the star where it has a hydrogen mass fraction less than 0.1, as was assumed in \citet{dewi00}. The term $L_{\star}$ is the luminosity of the star. We used $f=1$ otherwise. 

\subsection{Common Envelope Evolution}\label{subsec:ceevolution}

The planet can be engulfed by its host star if it is orbiting close enough during stages where the star increases in radius, such as the RGB or AGB. This event initiates a phase called Common Envelope evolution. During this process, the primary star fills its Roche lobe and wraps the companion planet into a common gas envelope. As the planet moves through this envelope, it experiences drag forces and begins to lose angular momentum and orbital energy. This usually causes a decrease in the planet's orbital separation. In the end, either enough orbital energy is injected into the envelope to eject it, and the planet remains in orbit, or the orbit continues to shrink until the planet is at or below its host star's convective zone. Then there can be two outcomes, either the planet is dissolved into the convective zone or it merges into the star's core \citep{Yarza2023}. In our simulations, we do not distinguish between these two destruction scenarios, we designate them both as ``accreted". 

The system will undergo common envelope evolution if 
$R_{\star}>R_{\mathrm{Roche}}$. This means that mass transfer begins when the star radius $R_{\star}$ reaches or exceeds its Roche radius $R_\mathrm{Roche}$ defined by the \citet{Eggleton83} approximation:
\begin{equation} \label{Roche}
    R_{\mathrm{Roche}}= a \frac{0.49q^{2/3}}{0.6q^{2/3}+\ln(1+q^{1/3})}
\end{equation}
where $q=M_{\star}/M_\mathrm{p}$.
After CE, the stripped core with mass $M_\mathrm{c}=M_\mathrm{WD}$ will be left. We define a new variable, $a_{\rm engulf}$, to describe the semi-major axis of the planet at the instant when Roche lobe overflow begins. For the CEE approach, we used the energy formalism by \citet{Webbink1984} combined with the numerical parameter $\alpha\lambda$ studied by \citet{deKool90}, where $\alpha$ is related to the CE efficiency and $\lambda$ to the central concentration of the body:

\begin{equation}
    \alpha \lambda \left(\frac{G M_\mathrm{c} M_\mathrm{p}}{2 a_f}- \frac{G M_{\star} M_\mathrm{p}}{2 a_{\rm engulf}}\right)= \frac{G M_\mathrm{env} M_{\star}}{R_{\mathrm{Roche}}}
    \label{eqn:CE}
\end{equation}
where $M_\mathrm{c}$ is the core mass of the host star, $M_\mathrm{env}$ is the mass of the stellar envelope, $M_{\star}$ is the total mass of the star, and $M_\mathrm{p}$ is the planet mass. We note that $\alpha$ is more specifically defined as the ratio of the energy needed to unbind the envelope to the orbital energy released during inspiral. In the simulations, this efficiency parameter was fixed at 0.41, 1, 2, 2.5, or 20 (we motivate such high values of $\alpha$ below).

We calculated $\lambda$ at each timestep by integrating the binding energy of the envelope:
\begin{equation}
    \begin{aligned}
        \lambda = -\frac{G M_{\star} M_\mathrm{env}}{a_{\rm engulf} r_L E_\mathrm{env}} \qquad
    \end{aligned}
    \label{eqn:lambda}
\end{equation}
where $r_L = R_\mathrm{Roche}/a_{\rm engulf}$ is the dimensionless Roche lobe radius of the star, so we can use $a_{\rm engulf} r_L = R_\mathrm{Roche}$. 
We defined $M_\mathrm{env}$ as the total mass above the star's core-envelope boundary, which is the location where the hydrogen mass fraction first exceeds 0.1 as defined in Sec \ref{sec_orbevol}). Here we calculate the binding energy $E_\mathrm{env}$ in a similar manner to Equation~5 of \citet{dewi00}, which included contributions of gravitational and internal thermal energy as:
\begin{equation}
    \begin{aligned}
        E_\mathrm{env} = -\int_{M_\mathrm{c}}^{M_{\star}} \frac{G M(r)}{r} \, dm + 
                \frac{6\pi k}{\mu \, m_{\mathrm{H}}}
            \int_{R_\mathrm{c}}^{R_{\star}} r^2 \, \rho(r) \, T(r) \, dr
    \end{aligned}
    \label{eqn:bind_energy}
\end{equation}
where $k$ is Boltzmann's constant, $\mu$ is the mean molecular weight $\simeq 0.615$, $m_H$ is the atomic mass of hydrogen, and $\rho(r)$ and $T(r)$ are the density and temperature radial profiles of the star, respectively.

\begin{figure}[!ht]
    \centering
	\includegraphics[width=1.0\columnwidth]{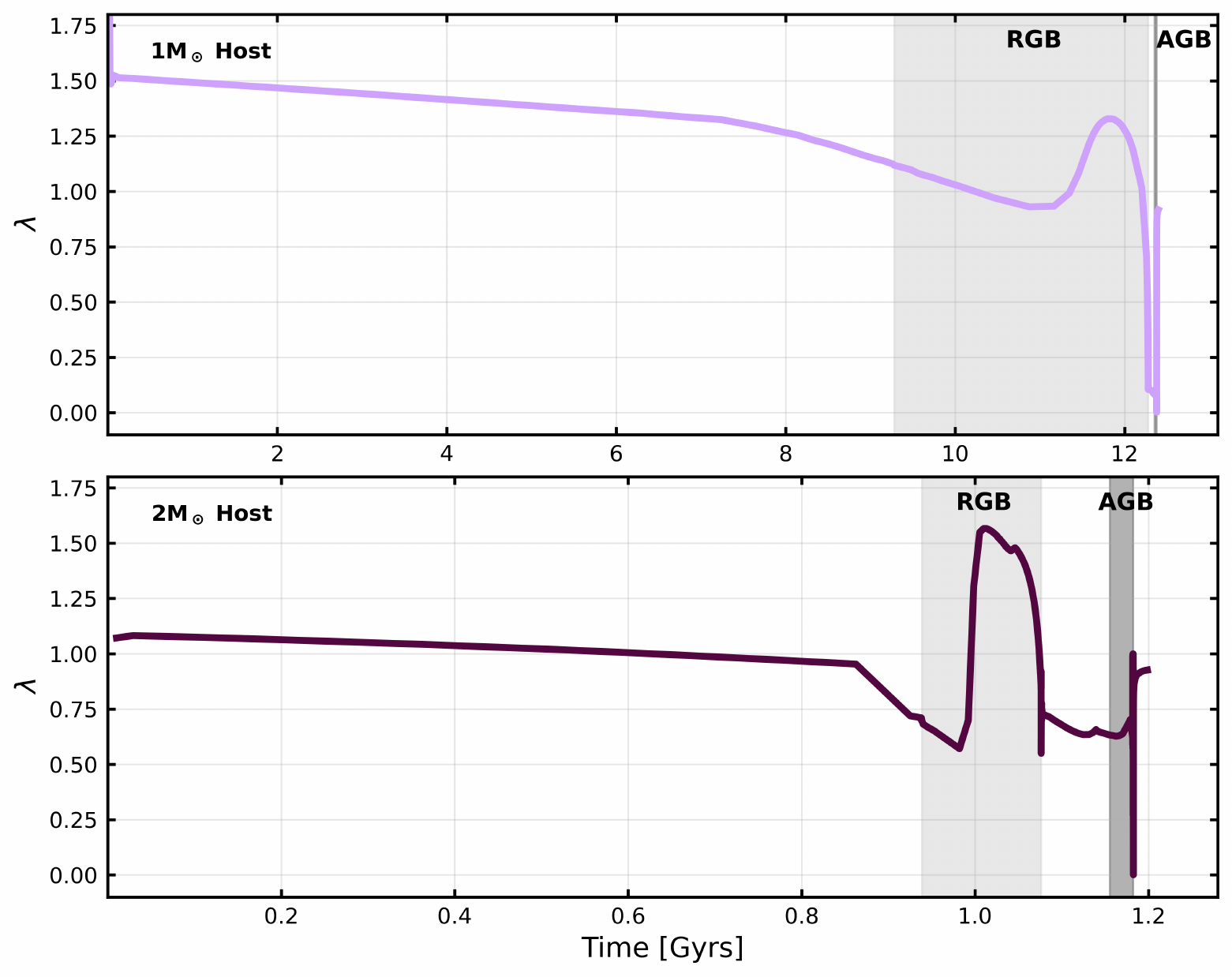}
	\caption{Time evolution of $\lambda$ at CE onset as calculated with Equation~\ref{eqn:lambda}. The upper panel shows the evolution for a $1\,M_{\odot}$ host, and the lower panel shows it for a $2\,M_{\odot}$ host. The extent of the red giant and asymptotic giant branches are indicated by the gray areas to highlight the changes in $\lambda$ as the host star evolves.
	\label{lambda_evol}}
\end{figure}

Note that we did not include ionization (recombination) energy in our calculations. This contribution can be comparable in magnitude to the thermal energy when the host star reaches larger radii and, if included, would drive the total envelope energy toward zero or even positive values if the ionization energy becomes sufficiently large:
\begin{equation}
    E_\mathrm{env} \approx E_\mathrm{grav}+E_\mathrm{th}+E_\mathrm{ion} \approx -V+V/2+V/2 \rightarrow 0 
\end{equation}
where $E_\mathrm{grav}$ is the gravitational potential energy with magnitude V and $E_\mathrm{th}$ is the thermal energy defined as with $E_\mathrm{th}=-E_\mathrm{grav}/2=V/2$ from the virial theorem. At large stellar radii, the ionization energy can approach or even exceed the thermal energy ($kT\sim$ binding energy per particle), which we express as $E_\mathrm{ion}\sim E_\mathrm{th}$. 
This would imply that no additional energy is required for envelope ejection. In this limit, the envelope becomes effectively unbound, and the parameter $\lambda$ would diverge to very large values (e.g. $\lambda\gtrsim$100). 
Given the uncertainty in how efficiently ionization energy contributes to unbinding the envelope, we decided to exclude it from our calculations of the envelope binding energy.
Previous work has also deemed it unnecessary to include recombination energy in population synthesis studies \citep{zorot22}.
However, to illustrate the potential impact of this additional energy, we included models assuming an envelope ejection efficiency parameter $\alpha$ greater than 1 (up to 20) in our simulations. 
Figure~\ref{lambda_evol} shows the time evolution of $\lambda$ for a $1$ and $2M_{\odot}$ host star.

The final semi-major axis $a_\mathrm{f}$ can then be solved for by substituting Equation~\ref{eqn:lambda} into Equation~\ref{eqn:CE}:

\begin{equation}
    a_\mathrm{f} = a_{\rm engulf}\frac{ M_\mathrm{c}}{M_{\star}} \left(1- \frac{2 a_{\rm engulf} E_\mathrm{env}}{\alpha G M_{\star} M_\mathrm{p}} \right)^{-1}
\end{equation}
However, because we already pre-calculated $\lambda$ at Roche overflow in our simulations, we substituted $\lambda$ explicitly into Equation~\ref{eqn:CE} after solving for $a_{\rm{f}}$ to track the orbit evolution:
\begin{equation} \label{eqn:fin_af_sim}
    a_\mathrm{f} = a_{\rm engulf} \frac{M_\mathrm{c}}{M_{\star}} \left( \frac{2 M_\mathrm{env}}{\alpha \lambda r_{L} M_\mathrm{p}} +1 \right)^{-1}
\end{equation}

\subsection{Termination conditions}

Not all planets survive during the CE phase of the simulations. We therefore defined planetary ``termination'' as the point at which a planet can no longer maintain a stable orbit or a temperature suitable for survival . Once the host star has filled its Roche lobe ($R_{\star}>R_\mathrm{Roche}$), we calculated $a_f$ for CE evolution. Then we assessed the potential destruction of the planet under two conditions. First, we considered whether a planet is tidally disrupted inside the star, and is either dissolved in the convective zone or completely accreted onto the stellar core. We check if this takes place by testing whether $a_\mathrm{f}$ is less than the planet's tidal disruption limit $a_{disrupt}$. We note that the tidal disruption radius for the planet is also known as the Roche limit, but to avoid confusion with the Roche radius of the star discussed in Section~\ref{subsec:ceevolution}, we will refer to this distance as the tidal disruption radius.  For Jupiter-like planets, we can solve for $a=a_{disrupt}$ from Equation~\ref{Roche} after setting $R_\mathrm{Roche}=R_\mathrm{p}$ and $q'=M_\mathrm{p}/M_\mathrm{WD}$:
\begin{equation}
    a_{disrupt}= \frac{R_\mathrm{p}}{0.49} \frac{0.6q'^{2/3}+\ln(1+q'^{1/3})}{q'^{2/3}}
\end{equation}
The second condition is when the planet ends up with a very short orbit, enough to be under the evaporation radius of the star $a_f<a_{evap}$ and become thermally destroyed. This is where the planet can experience significant irradiation from the stellar core during or after the envelope has been ejected. This distance was computed as in \citet{Soker98}: 

\begin{equation}
    a_{evap}=\left(\frac{M_p}{10\,M_{\rm{Jup}}}\right)^{-1} R_{\odot},
\end{equation}
where $M_p$ is the planet mass and $M_{\rm{Jup}}$ is Jupiter mass. Together, these criteria provide physical constraints for planetary survival after CE evolution.

\section{Results} \label{results}

\subsection{Possible Outcomes}
\begin{figure}[!t]
    \centering
	\includegraphics[width=1\columnwidth]{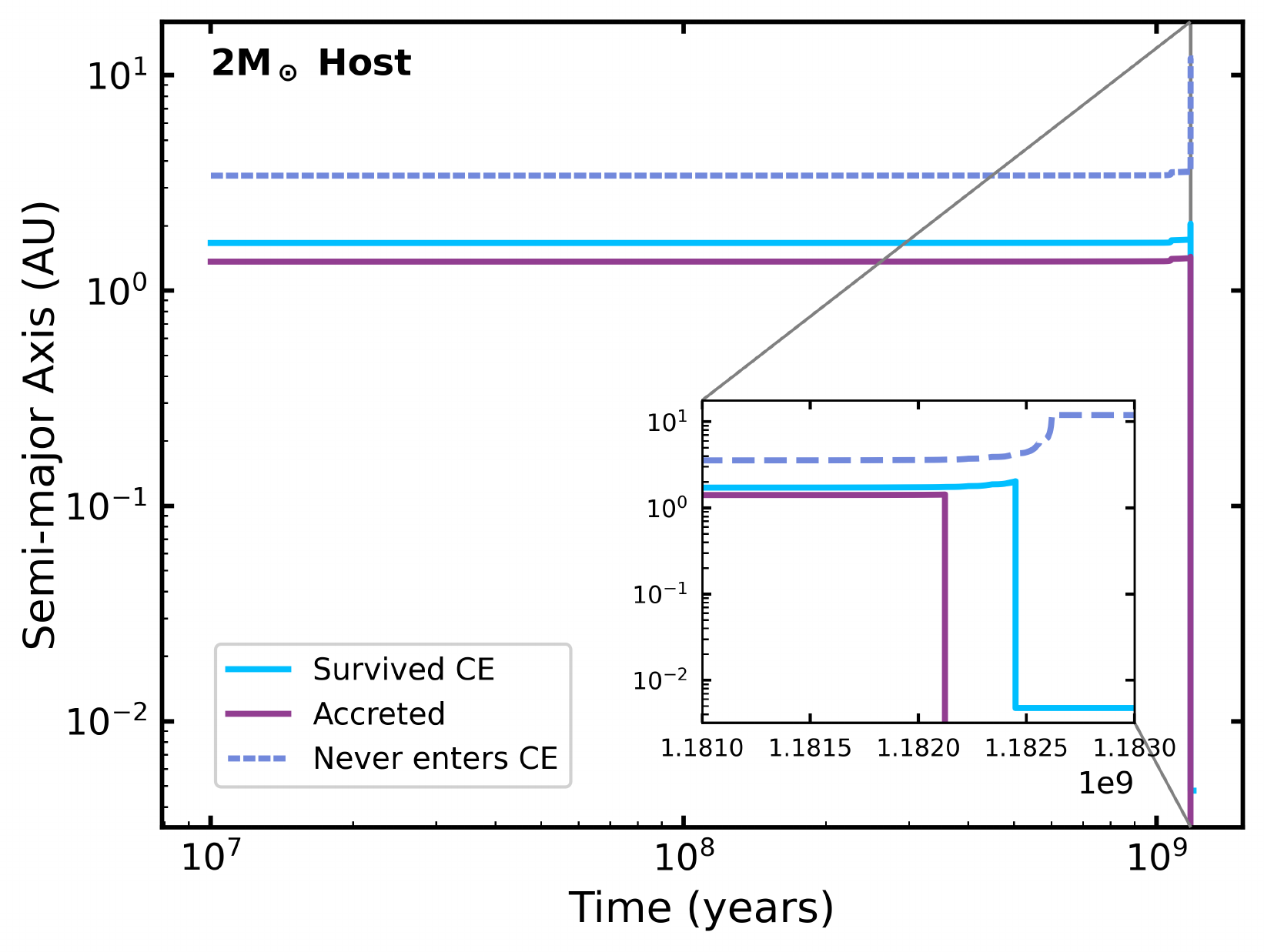}
	\caption{The planet semi-major axis evolution color-coded by scenario outcome for the case with static $Q'_\mathrm{{\star,p}}$ and $\alpha$=2. \newline Inset: Zoom-in of the same plot centered around the final stages of the planet. The surviving planet ends up on a very close orbit, while the accreted case is destroyed within the host star. In the case where the planet never enters CE, tidal evolution causes the planet to migrate into a wider orbit as the star loses mass at the end of the giant phase.
	\label{evolution}}
\end{figure}

Our simulations yield three distinct evolutionary outcomes for the planets in our modeled systems.
The first possibility is that the planet never enters CE evolution and its orbit remains wide and only evolving in response to stellar mass loss. This outcome is typically associated with planets that begin at large separations, well beyond the maximum radius of the host star during its AGB phase. In such cases, the planet avoids engulfment altogether, remaining gravitationally bound to the remnant white dwarf.
The second scenario is where the planet undergoes CE evolution and is destroyed because the orbital energy released during inspiral is insufficient to eject the envelope material and avoid either tidal disruption or evaporation. 
The third outcome is that the planet undergoes CE evolution and survives. We find that these survivor planets are generally massive and usually located slightly beyond their tidal disruption radii, where tidal disruption by the star and the planet's self-gravity are balanced. Although this is a rare outcome in our simulations, it is feasible for planets to survive CE evolution under specific assumptions (especially assuming high values of envelope ejection efficiency) and finely-tuned initial orbital conditions. 
Figure~\ref{evolution} illustrates the orbital evolution of systems representative of the three outcomes using an illustrative set of assumptions, which is detailed in Section~\ref{default_case}.

\subsection{Realistic Case: $2M_{\odot}$, variable $Q'_{\star,\rm{p}}$, $\alpha=0.41$}

We first performed a batch of simulations for the selected subset of the $10^7$ initial systems (as explained in Section~\ref{sec:initparams}) with what we consider to be the most realistic set of assumptions we could choose. We chose a stellar mass of 2 $M_{\odot}$ because this closely matches the most common progenitor for white dwarf stars in our galaxy. We used a tidal prescription that allows the tidal quality factor to change with the evolution of the host star, and we used a common envelope efficiency value that is in the upper range of plausible values of $\alpha$  by observations of the population of post-common-envelope white dwarf/brown dwarf binary stars \citep{zorot22}. Out of 10$^7$ initial systems in this instance of parameters, there were no systems where the planets survived CE evolution. This implies that if our assumptions indeed hold, planets surviving common envelope evolution are likely very rare, with an occurrence rate $\lesssim 10^{-7}$.

\subsection{Illustrative Case: $2M_{\odot}$, static $Q'_{\star,\rm{p}}$, $\alpha=2$} \label{default_case}
Because our most realistic set of simulation inputs and assumptions did not result in any surviving planetary systems, we also performed simulations with slight permutations to those realistic assumptions in order to illustrate the properties of the common envelope planet population. For this illustrative case, we continued to assume the planet orbited a $2 M_{\odot}$ star, and used a slightly simpler static tidal quality factor prescription with $Q'_{\star} = 10^5$ and $Q'_\mathrm{p} = 10^{5.5}$. Most importantly, we assumed a higher envelope ejection efficiency parameter $\alpha$ of 2.0 which implies some modest amount of additional energy has been added into the system, perhaps from recombination reactions. These parameters do result in a diversity of post-CE outcomes; a portion of the initial planetary distribution of our illustrative-case simulations is shown in Figure~\ref{color_initial}, with color coding corresponding to the planets' post-CE fates. After the cuts in parameter space made for computational efficiency (as described in Section \ref{sec:initparams}), we were left with 1,461,550 systems to run through our model. About $0.074\%$ of our simulated systems result in CE survivors, $18.27\%$ of planets are destroyed via evaporation, $1.7\%$ are destroyed via tidal disruption, and $\sim80\%$ of systems never go through CE.

\begin{figure*}[!htb]
    \centering
	\includegraphics[width=0.95\textwidth]{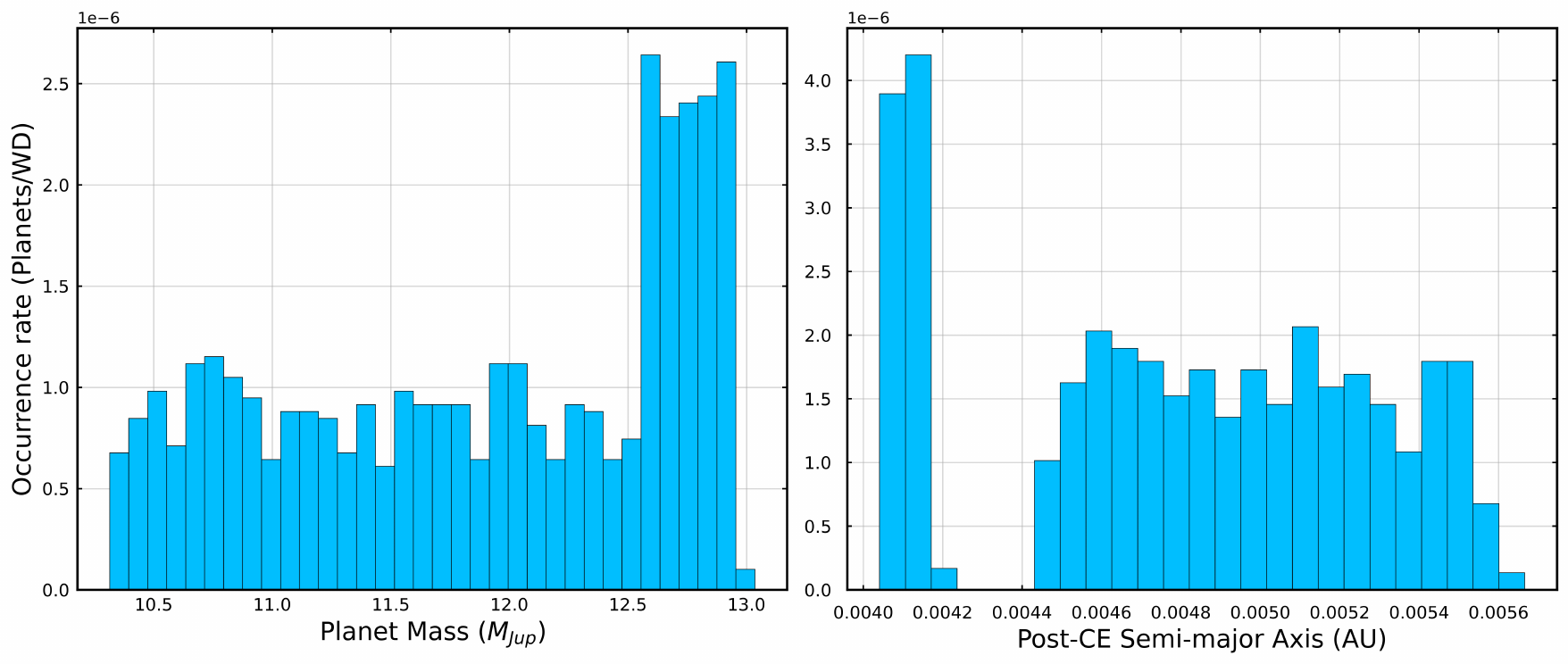}
	\caption{Final distributions of planets surviving CE evolution in the illustrative case (2\,$M_{\odot}$, static $Q'_{\star} = 10^5$ and $Q'_\mathrm{p} = 10^{5.5}$, $\alpha$=2). Our results show that planets surviving CE are massive and will evolve to very short orbits. The bimodality in semi-major axis and the spike in planet mass is due to planets surviving at two different points in stellar evolution corresponding to two different radii of their host star during the AGB stage.
	\label{default_summary}}
\end{figure*}

\begin{figure}[!htb]
    \centering
	\includegraphics[width=1\columnwidth]{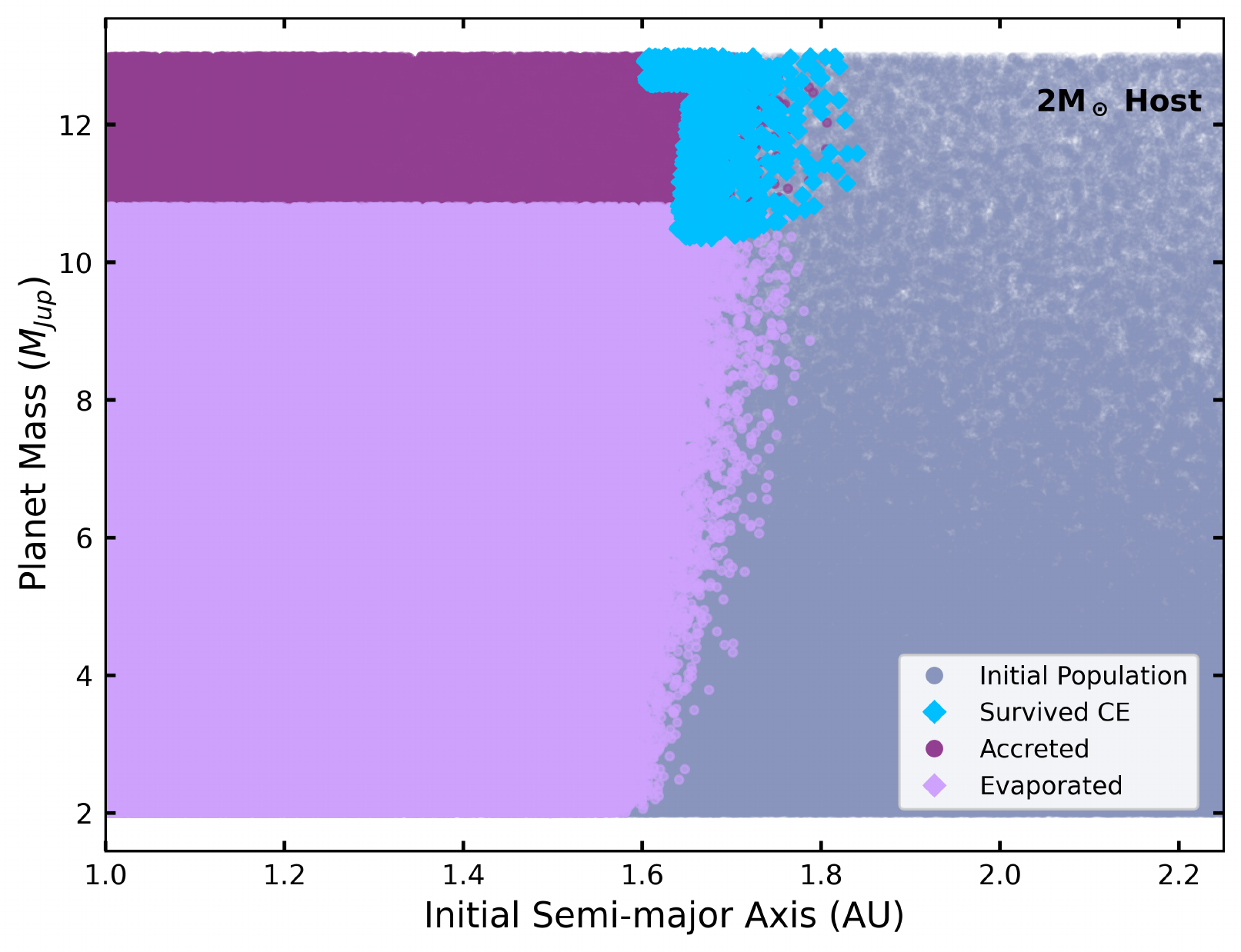}
	\caption{Initial planet mass vs semi-major axis of the population around a 2\,$M_{\odot}$ main-sequence host, run with static $Q'_{\star} = 10^5$ and $Q'_\mathrm{p} = 10^{5.5}$, $\alpha$=2 (illustrative case), and zoomed in around this area of interest. Each point is color-coded by outcome after the common envelope. Surviving planets are highlighted in light blue. 
    Planets that survive common envelope evolution tend to be massive ($\gtrsim 10 M_{\rm{Jup}}$) and orbit in a narrow range of semi-major axes near the boundary between planets that are tidally disrupted within the star and accreted onto the cores and planets that move outwards and never experience engulfment. 
	\label{color_initial}}
\end{figure}

Our simulations resulted in 1084 surviving planets, which is a small fraction ($\sim10^{-4}$) of the total 10 million initial systems. Even in this case with the additional energy to help the planet unbind the stellar envelope, CE survival is still a rare outcome. Importantly, because we have a sample of surviving planets, we can now examine the population characteristics of post-CE planets. We find that the surviving planets are usually massive ($\gtrsim 10 M_{\rm{Jup}}$) and orbiting very close to their star ($\lesssim 0.006$\,AU, $\sim 1.3 R_{\odot}$). The final distribution of the planets surviving CE in our illustrative case for a $2 M_\odot$ is shown in Figure~\ref{default_summary}.

\begin{figure}[!htb]
    \centering
    \includegraphics[width=1\columnwidth]{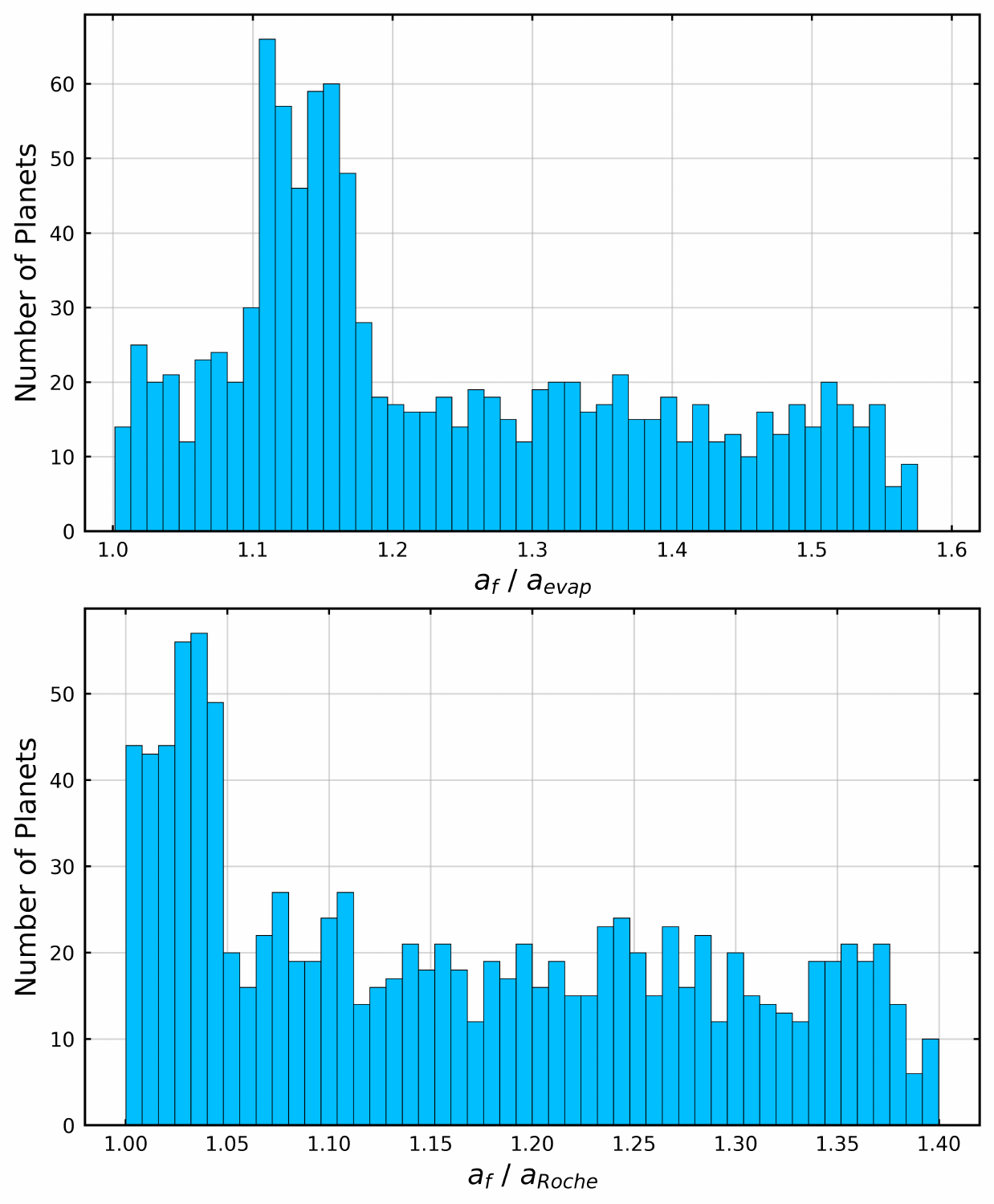}
	\caption{Final semi-major axis of the surviving planets in our illustrative case (2\,$M_{\odot}$, static $Q'_{\star} = 10^5$ and $Q'_\mathrm{p} = 10^{5.5}$, $\alpha$=2). The results are shown relative to the evaporation limit (top) and the tidal disruption limit (bottom). Planets that survive the CE phase typically have semi-major axes just beyond the evaporation or disruption limit (within a factor of 2 of each limit).
	\label{aroche}}
\end{figure}

We further investigated the distribution of the final semi-major axes for surviving planets by comparing them to the orbital radii at which they would either be tidally disrupted ($a_{\rm disrupt}$, also called the tidal disruption radius or the Roche limit) or the location within the star that they would have likely been evaporated ($a_{evap}$, which we call the evaporation radius). We show the ratio of the planets' final semi-major axes to these limits in Figure~\ref{aroche}.  We find that surviving planets pile up around as close an orbital separation as they can survive, either around the evaporation radius or the tidal disruption limit. All of the planets in this illustrative case had semi-major axes $a_f < 1.4~ a_{disrupt}$ and $a_f < 1.6 ~a_{evap}$.

\subsection{Varying Parameters}
In addition to running our simulations with these two specific sets of assumptions, we also systematically explored how the variations in key physical parameters (host star mass, tidal prescription, and common envelope efficiency) influence the final population of planetary systems that emerge from CE evolution. We ran a grid of simulations, each evolving the subset out of $10^7$ initial systems, over all possible combinations of two stellar mass values, three tidal prescriptions/quality factors, and five values of common envelope efficiency.  We summarize the number of surviving planets and their typical orbital parameters across our parameter space in Table~\ref{tab:result_sum}.

\begin{deluxetable*}{ccc|ccccc|ccccc}
\centering
\tablecaption{Number of CE survivors $N_{\rm surv}$ per 10 million systems for each host star mass and different $Q'_\mathrm{{\star,p}}$ and $\alpha$. We also include the median final semi-major axis ${a_f}_{\rm med}$ and planet mass ${M_{\rm p}}_{\rm med}$. We also present the occurrence rate and number of transiting survivors $N_{\rm{tr}}$ for each scenario.
\label{tab:result_sum}}
\tablehead{ \multicolumn{3}{c}{\textbf{Parameters}}  \vline & \multicolumn{5}{c}{\textbf{1M$_{\odot}$}}\vline & \multicolumn{5}{c}{\textbf{2M$_{\odot}$}} \\ \hline
\colhead{$Q'_{\star}$}& \colhead{$Q'_p$} & \colhead{$\alpha$} \vline & 
\colhead{$N_{\rm{surv}}$} & \colhead{\makecell{$a\rm{{_f}_{med}}$\\(AU)}} & \colhead{\makecell{$\rm{{M_p}_{med}}$ \\($\rm{M_{\rm{Jup}}}$)}} & \colhead{\makecell{Occ.\\Rate}} & \colhead{$N_{\rm{tr}}$} \vline & \colhead{$N_{\rm{surv}}$} & \colhead{\makecell{$a\rm{{_f}_{med}}$\\(AU)}} & \colhead{\makecell{$\rm{{M_p}_{med}}$ \\($\rm{M_{\rm{Jup}}}$)}} & \colhead{\makecell{Occ.\\Rate}} & \colhead{$N_{\rm{tr}}$}
}
\startdata
\multirow{4}{*}{variable} & \multirow{4}{*}{variable} & 0.41 & 0 & -& - &- &-           &0    &-       & -     & -  & -\\
                   &                           & 1          & 0  & - & - &- &-          & 142 & 0.0054 &11.17 & 9.9$\times 10^{-6}$& 10\\
                   &                           & 2          &0  & -& - &- &-            & 578 & 0.0075  & 10.93 & 4 $\times 10^{-5}$ & 53\\
                   &                           & 2.5         &737  & 0.0042& 12.63 &2.5$\times 10^{-5}$ & 85    &1046  & 0.0062& 11.17 & 7.3 $\times 10^{-5}$& 101\\
                   &                           & 20         &229907 & 0.0098& 9.54 &7.7$\times 10^{-3}$ & 14949   & 113407 & 0.0078 & 10.27 & 7.9$\times 10^{-3}$ & 8734\\ 
                   \cline{1-13}
\multirow{4}{*}{10$^5$} & \multirow{4}{*}{10$^{5.5}$} & 0.41  & 0  & - & - & - &-      & 0   &- & -  &- &-\\
                   &                                & 1       & 0  & - & -& -& -         & 0   &- & -  &- &- \\
                   &                                & 2      &0   & -& -& -& -           &1084 & 0.0049 &12.03 & 7.6$\times 10^{-5}$ & 147 \\
                   &                                & 2.5   &458  & 0.0042& 12.62 & 1.6$\times 10^{-5}$ & 79   &3845  & 0.0048& 11.98 & 2.7$\times 10^{-4}$ & 465 \\
                   &                                & 20    &147215 & 0.011 &9.15 & 4.9$\times 10^{-3}$ & 8473   &155628 &0.013 & 9.10 & 1.1$\times 10^{-2}$& 8991\\ \cline{1-13}
\multirow{4}{*}{10$^6$} & \multirow{4}{*}{10$^{5.5}$} & 0.41 & 0  & -& - & -&-   & 0  & -& -  & -& - \\
                   &                                & 1     & 0  & - & -&- & -    & 0  & -& - & -& -\\
                   &                                &2     &0  &  -& -&- & -      &989 & 0.0048& 12.15  & 6.9$\times 10^{-5}$& 126\\
                   &                                & 2.5  &598  & 0.0041& 12.67& 2$\times 10^{-5}$ & 86     &4072  & 0.0048& 12.11 & 2.9$\times 10^{-4}$& 533\\
                   &                                &20    &135472 & 0.012 & 8.91& 4.6$\times 10^{-3}$ & 7297  &197933 & 0.013 & 8.78 & 1.3$\times 10^{-2}$ & 10343 \\ \hline \hline
\enddata
\end{deluxetable*}

\subsubsection{Progenitor Mass}
\begin{figure}[!ht]
    \centering
	\includegraphics[width=0.45\textwidth]{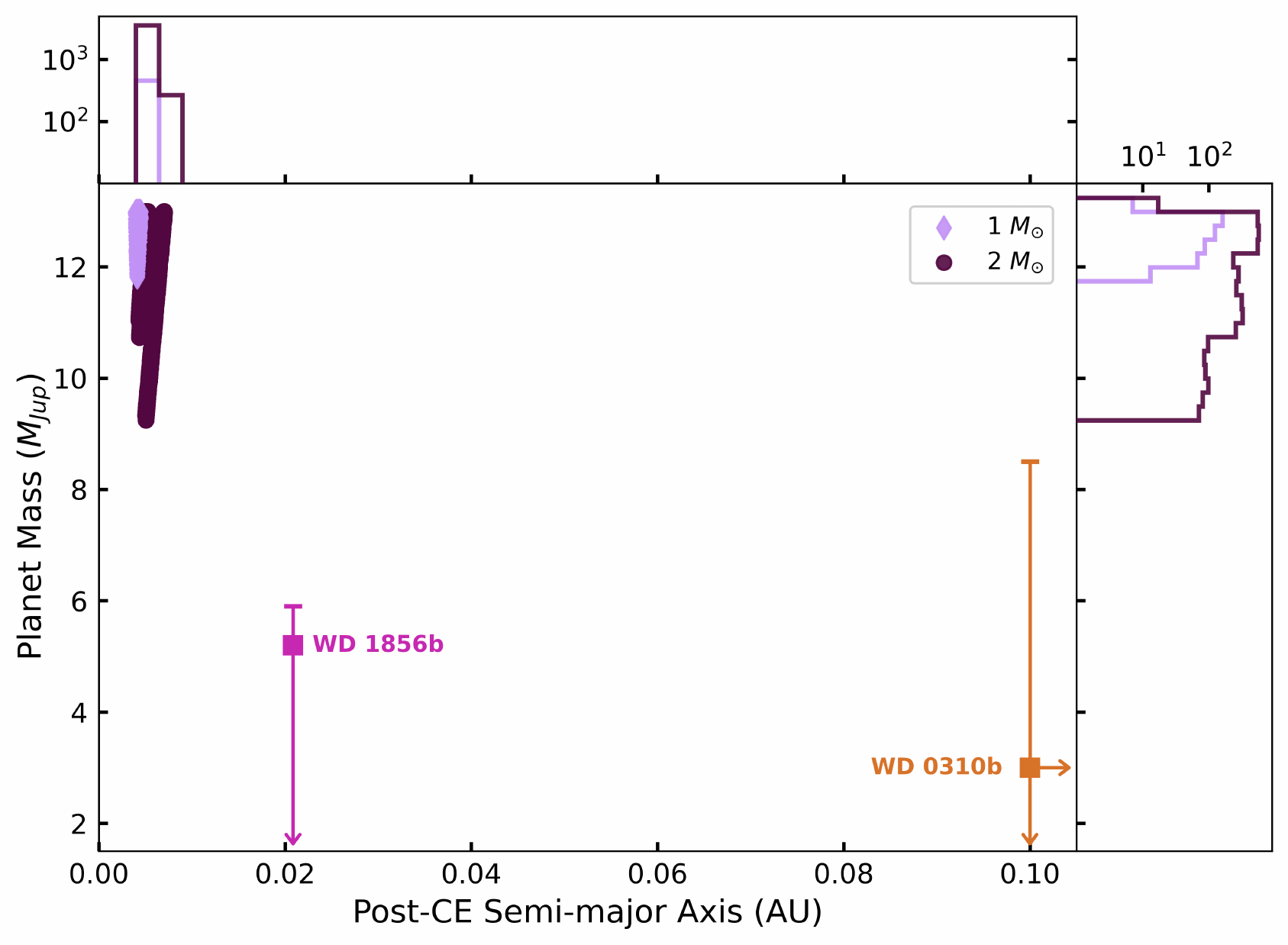}
	\caption{Joint distribution of planet mass vs final semi-major axis scatter plot shown for planets that survive CE around a 1 and 2\,$M_{\odot}$ main sequence hosts. We also show 1D histograms of planet mass and semi-major axis in the panels above and to the right of the main figure. The two sets of simulations we compare both used a static $Q'_{\star} = 10^5$ and $Q'_\mathrm{p} = 10^{5.5}$ and common envelope efficiency $\alpha=2.5$. We also plot the semi-major axis and mass of WD 1856b \citep{Vanderburg20} and the candidate WD 0310b \citep{limb24} in square markers for comparison. The inferred masses and orbital distances of these two objects fall well outside the typical distributions of planets surviving CE under these assumptions. 
	\label{progenitors}}
\end{figure}

We first investigated how the mass of the host star influences the resulting planet population. In addition to performing simulations with a $2 M_{\odot}$ progenitor (like in our illustrative and realistic cases), we also conducted simulations for a lower host mass of $1 M_{\odot}$ to explore how a less massive envelope and different evolutionary timescale affect the outcome. The comparison between the resulting planetary populations produced in these two cases is shown in Figure~\ref{progenitors} and Table~\ref{tab:result_sum}.

We find that overall in our simulations, there are roughly the same number of CE survivors around a host star of $1 M_{\odot}$ or $2 M_{\odot}$. In a case by case basis (that is, across different permutations of common envelope efficiency and tidal prescriptions), it is often but not always true that there are fewer CE survivors for a $1 M_{\odot}$ host than a $2 M_\odot$ host. This result is slightly counterintuitive because a $1 M_{\odot}$ star has a lower-mass envelope, which would (all else being equal) be easier to eject, allowing more planets to survive. 
However, even though the mass of the envelope is lower for a $1 M_{\odot}$ star, it turns out that $\lambda$ is lower, implying the envelope is more centrally concentrated and more strongly bound to the star, making it harder to eject.
The mean $\lambda$ at onset for CE survivors around a $1 M_{\odot}$ host is 0.18, while for the survivors around a $2 M_{\odot}$ this is 0.97. The larger values of $\lambda$ mean that less energy from the planet is required to expel the envelope, therefore allowing a larger number of CE survivors.

Overall, regardless of the mass of the host star, we found that the population of post-CE planets are qualitatively similar to our illustrative case. In both host-star scenarios, only the most massive planets survive, and they end up on very small final semi-major axes. 

\begin{figure*}
    \centering
    \includegraphics[width=0.95\textwidth]{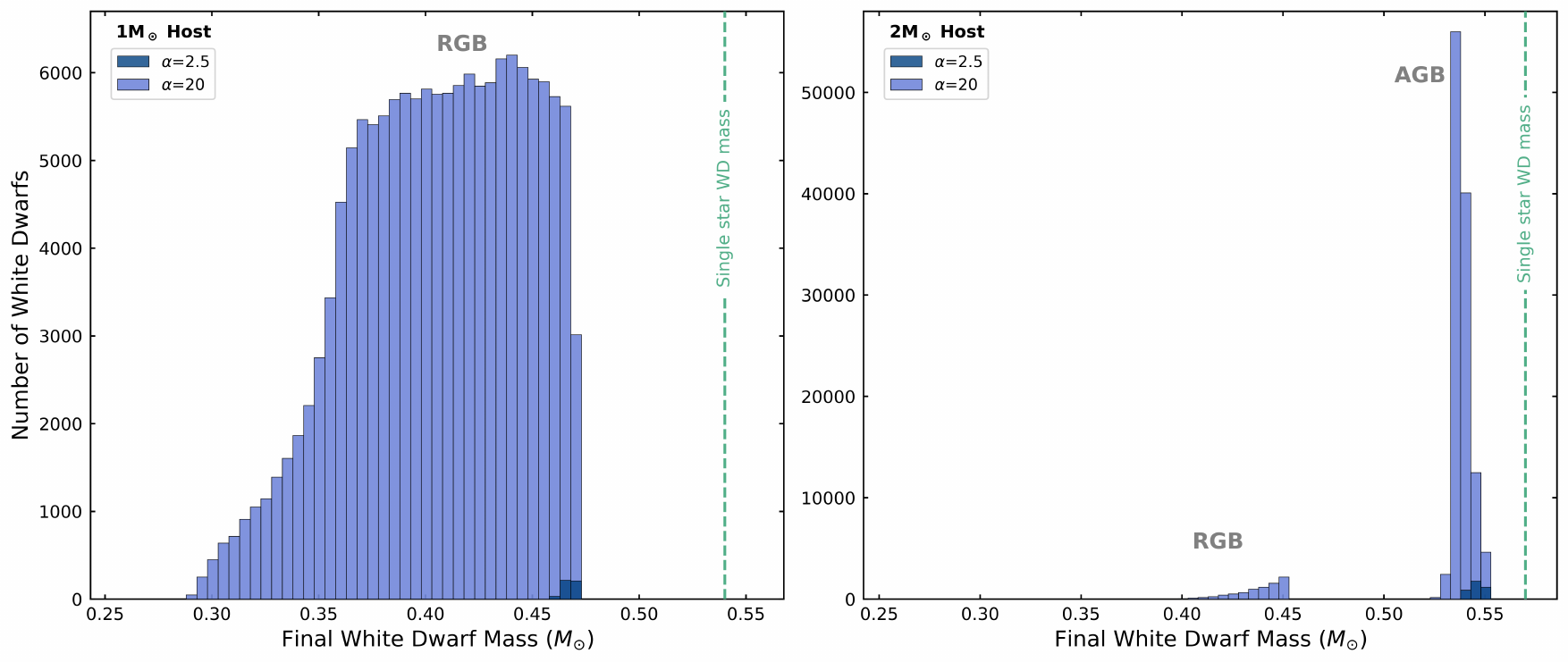}
    \caption{Final white dwarf mass distributions of systems with CE survivors around a 1 $M_{\odot}$ (left) and 2 $M_{\odot}$ (right) progenitors with a static $Q'_{\star} = 10^5$ and $Q'_\mathrm{p} = 10^{5.5}$ and color-coded by for two values of $\alpha$. We mark (green dashed lines) the typical mass of a WD that would result from each progenitor mass if the system did not go through CE. We also note the stellar evolution phase that corresponds to each feature in the distributions. For the 1 $M_{\odot}$ host, CE occurs during the RGB phase and leaves behind less massive WDs. For the 2 $M_{\odot}$ host, CE happens mostly during the AGB phase and result in slightly more massive WDs (with the exception of the $\alpha=20$ case which can have survivors during the RGB phase as well).}
    \label{fig:wdmass}
\end{figure*}

In scenarios where a planet does not undergo CE, stellar evolution will lead to a 0.57$M_{\odot}$ WD for a  $2 M_{\odot}$ progenitor and 0.54$M_{\odot}$ WD for a $1 M_{\odot}$ progenitor. However, when a planet and star undergo CE, the resulting WD mass spans a range of values set by the evolutionary phase at which CE occurs. Core mass grows continuously as the star ascends the giant branch. Therefore, earlier CE episodes (such as during the RGB) will result in lower-mass white dwarfs while later ones will leave slightly more massive remnants. This effect produces a spread in final masses of WDs hosting CE survivors, rather than the single mass value expected for systems that never undergo CE. Figure ~\ref{fig:wdmass} shows the final WD mass distributions for the systems that survive CE around a $1 M_{\odot}$ and $2 M_{\odot}$ progenitor with different values of $\alpha$. For the $1 M_{\odot}$ host, white dwarf masses are consistent with CE occurring during the RGB phase. The $\alpha=20$ case produces a broader distribution that includes lower masses ($\sim$ 0.3 $M_{\odot}$), while the $\alpha=2.5$ case has WD masses clustered closer to the upper end of the range ($\sim$ 0.45 $M_{\odot}$).
For the $2 M_{\odot}$ host, there can be two distinct WD mass distributions that can arise. The $\alpha=20$ case produces survivors from both the RGB ($\sim$ 0.4 $M_{\odot}$) and AGB ($\sim$ 0.54 $M_{\odot}$) stages. In contrast, the $\alpha=2.5$ case only produces AGB survivors clustered toward the upper end of the mass range.

\subsubsection{Tidal Prescriptions}
\begin{figure}[!t]
    \centering
	\includegraphics[width=0.45\textwidth]{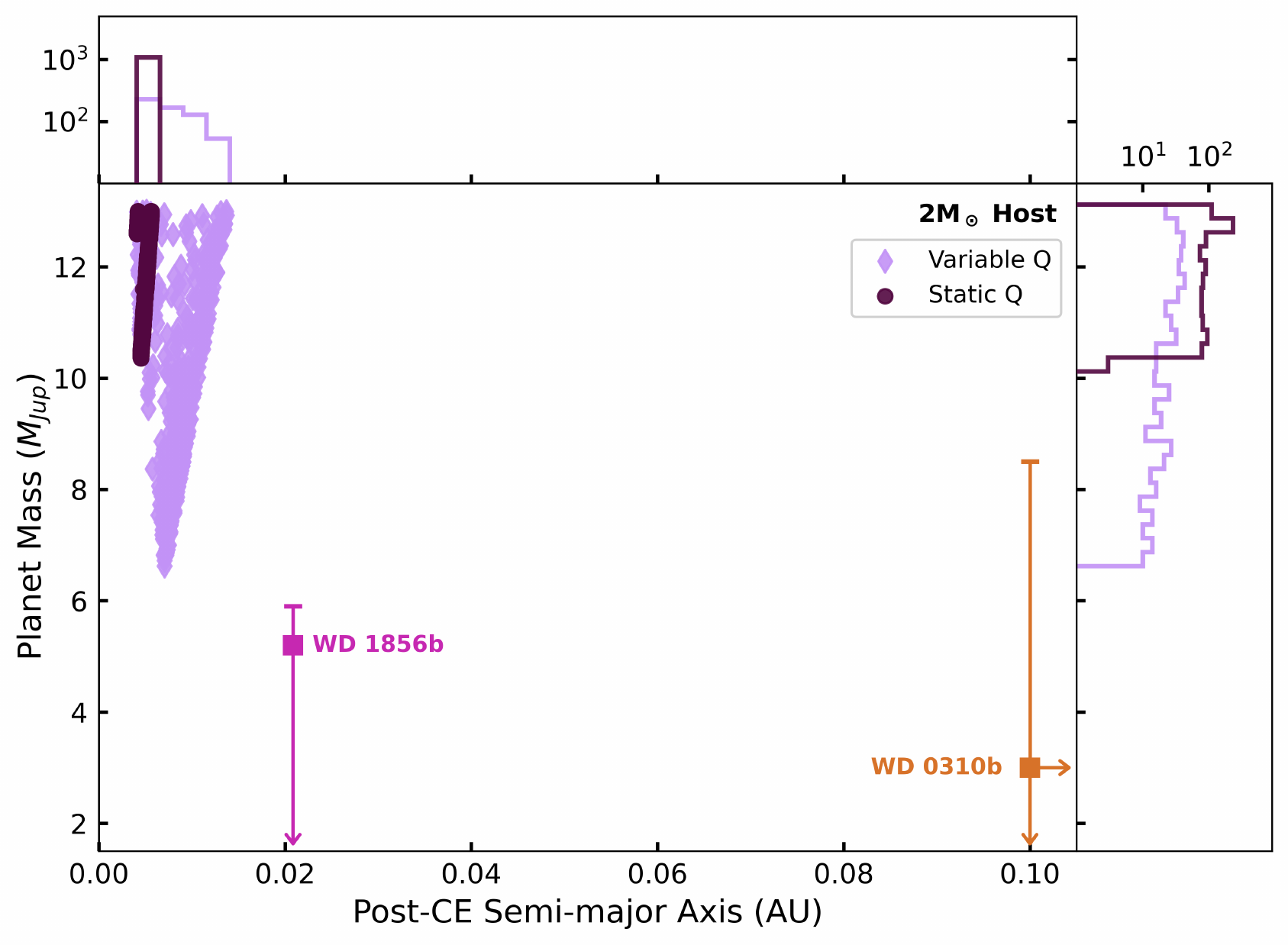}
	\caption{Joint distribution of planet mass vs final semi-major axis scatter plot shown for both the static ($Q'_{\star} = 10^5$ and $Q'_\mathrm{p} = 10^{5.5}$) and variable $Q'_\mathrm{{\star,p}}$ tidal prescriptions. We also show 1D histograms of planet mass and semi-major axis in the panels above and to the right of the main figure. Both sets of simulations assume a 2\,$M_{\odot}$ main sequence host and common envelope efficiency $\alpha=2$. In this case, we find that while variable tidal quality factor has fewer survivors overall, lower-mass planets can survive. Once again, we overplot the masses and semi-major axes for WD 1856b and WD 0310b for comparison.
	\label{tidal_prescrip}}
\end{figure}

We next tested how different tidal prescriptions affect the results of our simulations. We compared two tidal prescriptions to assess the role of tidal physics in shaping the post-CE planetary populations. One prescription assumed constant tidal quality factors for both the host star $Q'_{\star}$ and the planet $Q'_p$, while the other allowed them to vary. Variable tidal quality factors may be more realistic because tidal dissipation efficiency is tied to the orbital decay rate and can evolve meaningfully over the course of the stellar lifetime.  In the static case, we tested two values for $Q'_{\star}$: $10^5$ and $10^6$ as explained in Section~\ref{sec_orbevol}. 
The results of these tests are summarized in Table~\ref{tab:result_sum}, and a comparison of the populations for a particular host star mass ($2 M_{\odot}$) and common envelope efficiency ($\alpha = 2$) are shown in Figure~\ref{tidal_prescrip}. 

We found that generally, using different tidal prescriptions and different static tidal quality factors does change the number of surviving post-CE planets, but the changes are usually small, and there are no clear rules or patterns to predict whether one prescription will produce more CE survivors than another. An exception to this trend is in the limit of low common envelope efficiency, and cases where the planet has relatively low amounts of orbital energy (planets with relatively low-masses and large $a_f$). In these cases, it appears that a variable $Q'_{\star, p}$ prescription increases the likelihood that low-energy planets survive CE evolution. This is visible in Figure~\ref{tidal_prescrip}, where the surviving planets in the variable $Q'_{\star, p}$ prescription have masses as low as 7\,$M_{\rm{Jup}}$, while planets surviving in the static $Q'_{\star, p}$ case all have masses greater than 10\,$M_{\rm{Jup}}$. 

Overall, however, we continue to see the same qualitative picture of the post-CE planet population regardless of the tidal prescription: planets that survive CE are massive, and have small semi-major axes. 

\subsubsection{Efficiency Parameter}
Finally, we examined how variations in the envelope ejection efficiency parameter, $\alpha$, influence the survival of planets after CE. This parameter encapsulates the efficiency with which orbital energy is converted into kinetic energy to eject the envelope. 
In our simulations, we kept $\alpha$ constant in each run and varied the test value from 0.41 to 20 to capture conservative and optimistic scenarios. 
We include $\alpha=0.41$ as the upper limit within the range of 0.2--0.41 estimated from population-wide observations of post-CE WD binaries \citep{zorot22,zorot10, toon13, cama14}. We note that while some studies incorporate some internal energy sources into their calculations of $\lambda$ (specifically \citealt{zorot10} and \citealt{cama14}), which makes direct comparisons of inferred $\alpha$ challenging, the value we adopt of $\alpha=$0.41 inferred by \citet{zorot22} assumes no additional energy sources in their calculation of $\lambda$. 

\begin{figure}[!t]
    \centering
	\includegraphics[width=0.45\textwidth]{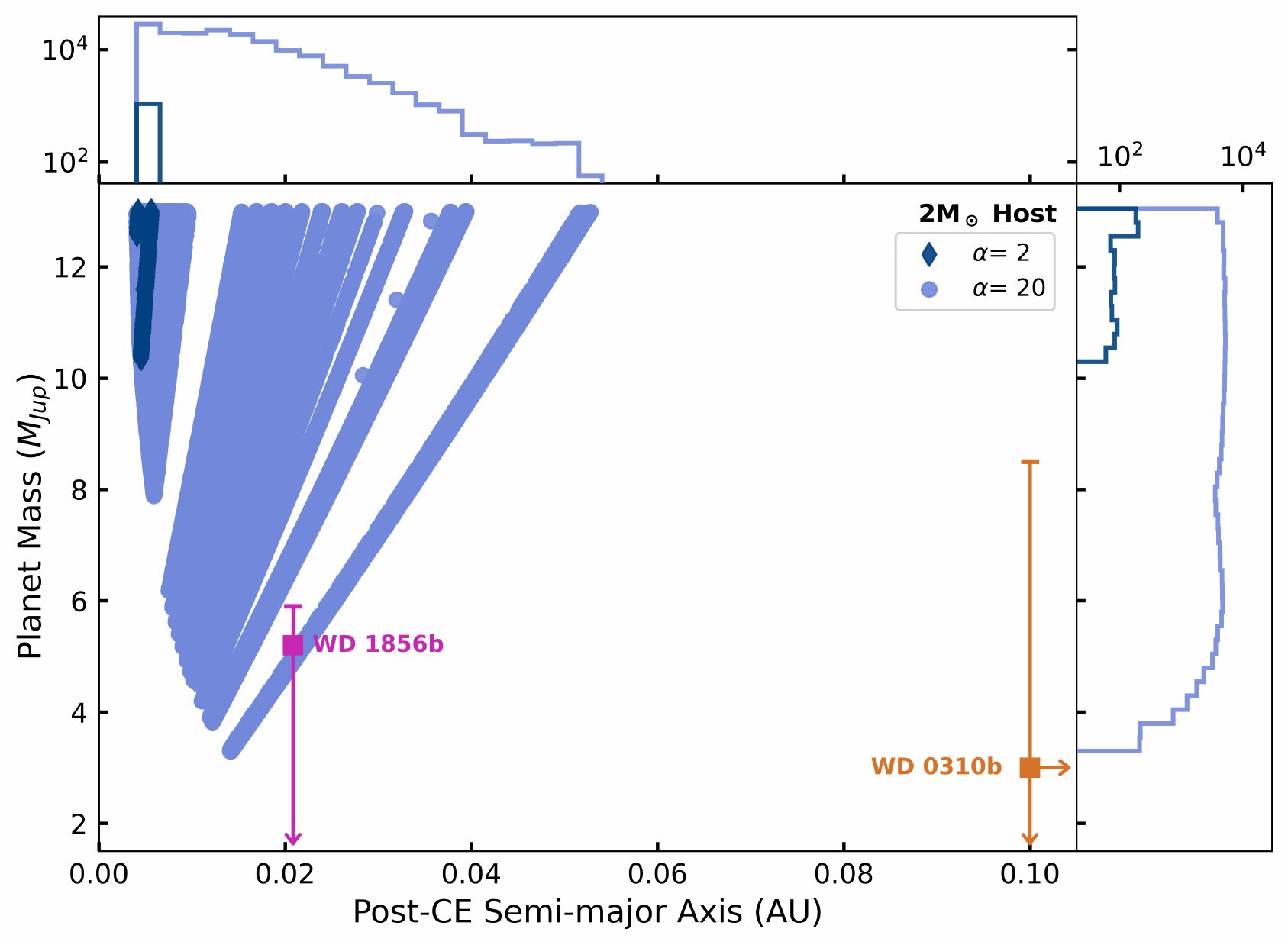}
	\caption{Joint distribution of planet mass vs final semi-major axis scatter plot shown for different values of $\alpha$ (2 and 20) . Final planet mass vs final semi-major axis scatter plot for planets that survive CE, shown for efficiency parameters $\alpha$=2 and 20. This is for planets around a 2\,$M_{\odot}$ main sequence host. This instance was run with a static $Q'_\mathrm{{\star,p}}$ prescription. The case with CE efficiency $\alpha=$20 is the only simulation we ran that reproduces planets with WD 1856b’s mass and semi-major axis; however, it also produces a large population of planets with smaller semi-major axis and higher mass that are more prevalent and would be easier to detect. Furthermore, such values of $\alpha$ are extreme and may not be physically realistic. The different lines are a result of planets surviving along different peaks in the RGB and AGB.
	\label{alphalambd}}
\end{figure}

The results of these experiments are summarized in Table~\ref{tab:result_sum}, and a comparison of the population for a particular tidal prescription and stellar mass is shown in Figure~\ref{alphalambd}. We find that modifying the common envelope efficiency parameter $\alpha$ causes by far the largest and most dramatic changes to the post-CE planet population. Increasing $\alpha$ from 0.41 to 20 increases the number of surviving planets by more than 5 orders of magnitude. 
While we inferred that post-CE planets are likely rare based on the results of our illustrative and most realistic cases, which yielded survival fractions of $\sim 10^{-5}$ and $\sim 10^{-7}$ respectively, if a considerable amount of additional energy is available to help unbind the envelope, the survival rates could be as high as $\sim 10^{-2}$. 

Like in our other tests, the population of surviving planets still skews towards massive planets and close-in orbits. However, this preference for massive planets is not surprising. We have derived an analytical expression (see Appendix) for the minimum planet mass threshold needed to survive CE: $\alpha \lambda M_p \gtrsim 20 \, M_{\rm Jup}$. For the $\alpha = 20$ case, it is possible for planets to survive with lower-masses and larger semi-major axes than in any of our other simulations.

While this high value of $\alpha$ is not needed to explain the population of stellar and brown dwarf companions in our galaxy \citep{zorot22}, it remains unclear whether the same efficiencies apply in planet-star interactions, so it is not out of the question that a considerable amount of additional energy can be injected in planetary-mass CE events.

\section{Discussion} \label{discuss}

\subsection{Occurrence Rates of CE survivors}
Our Monte Carlo approach allows us to predict the population of CE planets, rather than just the analytical parameters needed for a planet to survive CE. In many of our simulations, we noted that the fraction of planets that undergo and survive CE evolution is small. Here, we quantify this result by calculating the predicted occurrence rates of post-CE planets around white dwarfs.

Since every white dwarf originates from a main-sequence progenitor, we can calculate the predicted occurrence rate of common envelope survivors by multiplying the fraction of surviving planets in our simulations ($N_\mathrm{surv}/10^{7}$) by the occurrence rate of planets around main sequence stars. We source the occurrence rate of giant planets around main sequence stars (0.339 planets per star) from \citet{Fulton2021}, which was calculated for planets in a sample of stars with masses similar to the Sun. We also account for the dependence on host star mass \citep{Ghezzi2018ApJ}, so our final distributions are scaled by:
\begin{equation}
    f=0.339\left(\frac{M_{\star}}{M_{\odot}}\right)^{1.05}
\end{equation}
where $M_{\star}$ is the host star mass of 1 or 2$M_{\odot}$.
We report the results of this calculation, the occurrence rate of post-CE planets per white dwarf, in Table~\ref{tab:result_sum}.

The number of survivors primarily depends on the value of the common envelope efficiency $\alpha$ that we use. For the observationally-motivated value of $\alpha=0.41$ around a $2\,M_{\odot}$ WD progenitor, our simulations predict that we should find almost no CE-survivors; which indicates an upper limit of one survivor for every $10^{7}$ white dwarfs. For $\alpha=$2, we should find roughly one CE-survivor planet for every $10^4$ white dwarfs. In the scenario in which we tested the most additional energy injected, $\alpha=$20, we predict one planetary CE survivor out of every $10^2$ white dwarfs.

\begin{figure*}[!t]
    \centering
	\includegraphics[width=\textwidth]{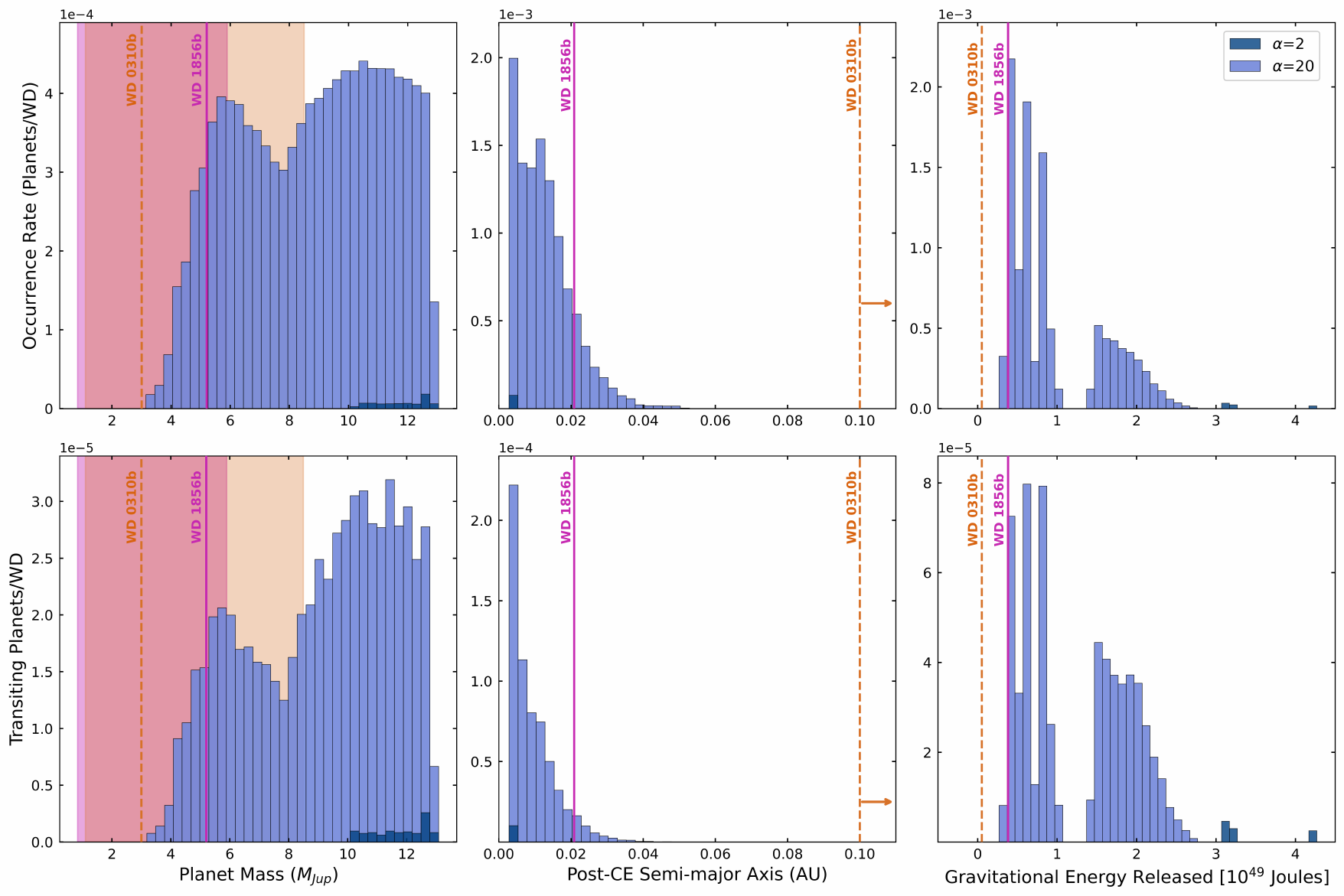}
	\caption{Final distributions of CE survivors around a 2\,$M_\odot$ WD progenitor. The histograms show planet mass (left), semi-major axis (center), and maximum gravitational energy released during CE (right) for two values of $\alpha$. The top panels show occurrence rates in the number of planets per WD. The bottom panels show the number of transiting planets per WD. We plot the mass and orbital positions of WD 1856b (solid magenta) and WD 0310b (dotted orange) with vertical lines. Shaded regions show the mass uncertainties for each one.
	\label{occrates_transit_summary}}
\end{figure*}

\subsection{Comparison to Known Planets}
One of the key motivations for this work was to shed light on how planets like WD 1856b and the candidate WD 0310b come to arrive in their present-day orbits\footnote{We do not include the candidate orbiting WD J091405.30+191412.25 in our comparisons given its highly uncertain mass (which could be similar to Neptune, or as high as 30\,$M_{\rm{Jup}}$, a factor of $\sim 10^3$) \citep{Gansicke19}.}. One way we can better understand the origin of these planets is to compare their properties (in particular, mass and orbital separation) to theoretical predictions of planets that undergo various formation pathways. Here, we review the properties of WD 1856b and WD 0310b and compare them to the outcomes of our CE simulations.

WD 1856b orbits its white dwarf host every 1.4 days, and has an orbital separation of about 0.02\,AU \citep{Vanderburg20}. The mass of WD 1856b is likely about 5 Jupiter masses ($M_{\rm p}=5.2^{+0.7}_{-0.8}\,M_{\rm{Jup}}$), but could be as low as 0.84\,$M_{\rm{Jup}}$ \citep{Xu2021AJ, limbach25}. Meanwhile, if real, the candidate planet WD 0310b has a semi-major axis between $0.1$ and $2$\,AU, and a mass between 0.2 and 3$\,M_{\rm{Jup}}$.

Immediately, the masses and orbits of these two planets stand apart from the population of CE survivors predicted in this work. In our simulations, we found that typical survivors of common envelope evolution have masses $\gtrsim 10\,M_{\rm{Jup}}$ and semi-major axes $\lesssim 0.01$\,AU. Both WD 1856b and WD 0310b are less massive and have larger orbital separations than the typical common envelope survivor. The physical reason why common envelope survivors are dominated by closely-orbiting and massive planets is because these planets can release the most gravitational potential energy to unbind the stellar envelope. We therefore compared the mass, semi-major axis, and maximum amount of gravitational potential energy released ($U_{max} = GM_c M_p/a_f$) for the known planets and our simulation results in Figure~\ref{occrates_transit_summary}. 

\begin{figure*}
    \centering
    \includegraphics[width=0.95\textwidth]{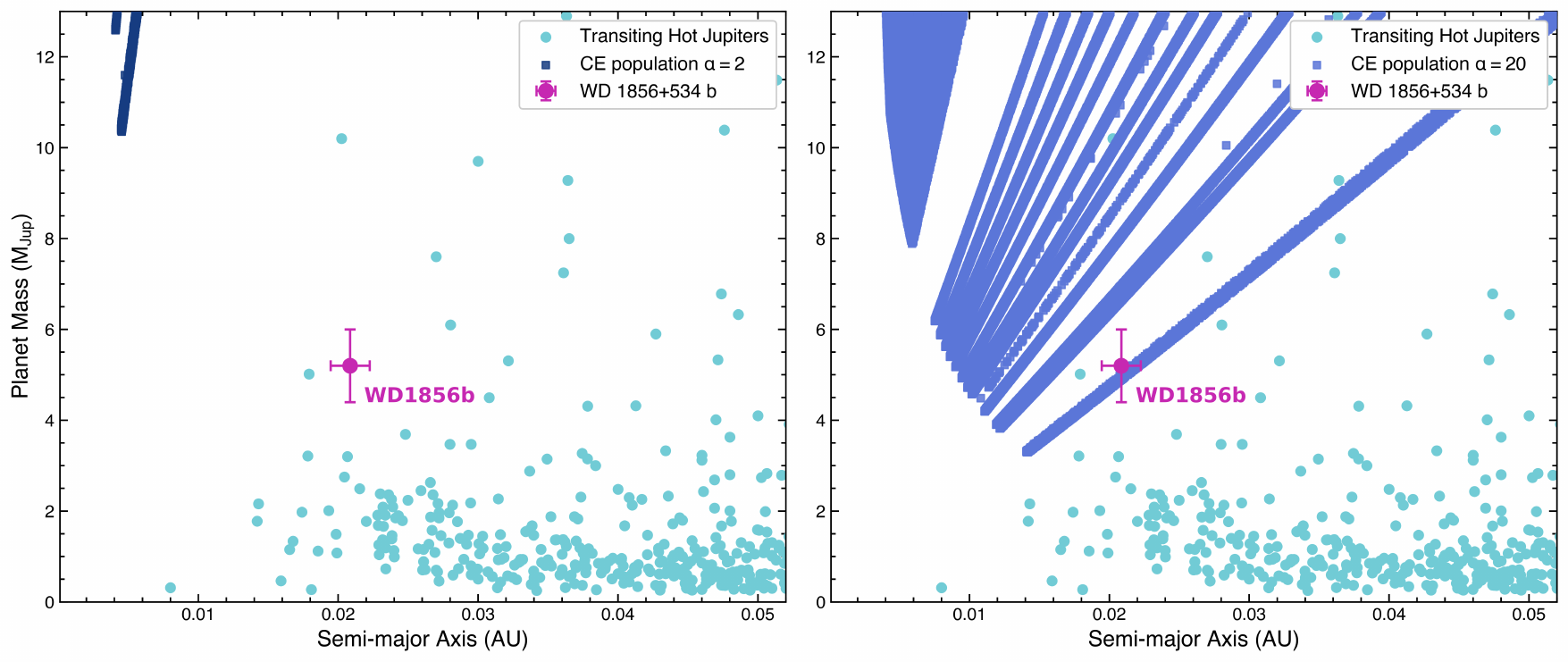}
    \caption{Planet masses and semi-major axes of CE survivors around a 2$M_{\odot}$ WD progenitor run with a static $Q'_\mathrm{{\star,p}}$ prescription (squares) and the transiting hot Jupiter population (circles). WD 1856b is also marked in pink with error bars. CE survivors with $\alpha=2$ (left) do not coincide with the properties of WD 1856b. Instead, this WD planet mostly agrees with the transiting hot Jupiter population, unless we consider cases with a higher CE efficiency such as $\alpha=20$ (right), which can encompass planets like it.}
    \label{fig:MS_alphas}
\end{figure*}

The known closely-orbiting white dwarf planets are highly discrepant from the results of all of our simulations. None of our simulations produce planets with as large semi-major axes and low masses as WD 0310b, and only the most extreme $\alpha=20$ case with the most additional energy used to eject the stellar envelope can match the parameters of WD 1856b. Even in this case WD 1856b occupies the lowest-energy tail of the distribution of common envelope survivors. If common envelope evolution were operating efficiently enough to form these two planets, it seems unlikely that two early discoveries would occupy such sparsely-populated regions of parameter space. Instead, it seems more likely that another formation mechanism is responsible for these objects. WD 1856b's orbit, in particular, is highly consistent with the population of hot Jupiters around main sequence stars \citep[see Figure 10 of][]{Xu2021AJ}. We show this in the context of transiting hot Jupiters (planet mass between 0.25-13$M_{\rm{Jup}}$ and orbital period $<$10 days) and our CE population around a 2$M_{\odot}$ main sequence host run with different values of $\alpha$ in Figure \ref{fig:MS_alphas}.

We can also test whether the occurrence rate for CE survivors we predict is compatible with the presence of WD 1856b and WD 0310b in the solar neighborhood. Both systems are extremely close to the Sun in cosmic terms. WD 1856 is among the 250 closest white dwarfs within 40\,pc of the Sun, while WD 0310 ranks among the 25 closest \citep{Obrien24}. We found that in most of our simulations, CE survivors are rare; our most realistic simulations predict occurrence rates less than $10^{-7}$. It is therefore very unlikely that two examples of such a rare process would be found so nearby.

\subsubsection{Population of Transiting Survivors}

The populations and occurrence rates we have predicted in this work do not account for the probability that the objects will be detected in observations. Because WD 1856b, the most confident example of a closely orbiting planet around a white dwarf, was discovered via transits, we simulate how the requirement that detected planets transit their white dwarfs affects our predicted population. 

To correct for this bias, we can select systems based on their transit probability:
\begin{equation}
    P_{tr}=\frac{R_{\rm{p}}+R_{WD}}{a_f}
\end{equation}
where $R_{\rm{p}}$ is a planet radius, $R_{WD}$ is the radius of the white dwarf in our simulations (0.015\,$R_{\odot}$ for a 2\,$M_{\odot}$ progenitor), and $a$ is the final semi-major axis of each planet. 
For each planet, we draw a random number from a uniform distribution between 0 and 1. If the random draw is $\leq P_{tr}$, then we include the planet in the transiting sample.
We summarize and compare the final distributions of transiting CE survivors with different $\alpha$ values in the bottom panels of Figure~\ref{occrates_transit_summary}. We collect the number of transiting planets per white dwarf for each of our simulation runs and present them in Table~\ref{tab:result_sum}. 

We find that the outcome of this calculation is that the population of \textit{transiting} CE survivor planets is even more strongly biased towards small semi-major axes, making the relatively wide orbits of WD 1856b and WD 0310b stand out even more. Another consequence, as expected, is that only a small fraction of planets transit their stars, so transiting CE survivors are very rare. We predict that the fraction of white dwarfs with transiting post-CE planets is $\sim10^{-5}$. This fraction is small enough that existing observations might not be expected to detect any such objects. For example, one of the main observatories that has been used to search for transiting planets around white dwarfs (and that discovered WD 1856b) is the Transiting Exoplanet Survey Satellite (\textit{TESS}, \citealt{Ricker15}), which has looked at an order of $\sim10^{4}$ WDs. If the fraction of white dwarfs with transiting CE survivors is really as low as our simulations predict, TESS has not looked at enough white dwarfs to likely detect such a system.

Fortunately, other datasets exist that provide an opportunity to detect CE survivors. In particular, there are ongoing efforts (e.g. Householder in prep.) to search through ground-based surveys like the Zwicky Transient Facility (ZTF, \citealt{2019PASP..131a8002B}). Rich datasets like ZTF or the future Legacy Survey of Space and Time (LSST) conducted by the Vera C. Rubin Observatory \citep{Ivezic19}, which have and will observe billions of objects, present an opportunity to detect very rare objects like post-CE planetary systems.

\subsection{Future Work}

This work is the first attempt to model a full post-CE planet population. However, there is more physical complexity that can be implemented into our approach to build a more complete picture of CE evolution and its outcomes for the formation of WD-planet systems.
The first avenue for improvement is to consider alternative CE outcomes beyond the canonical ejection scenario. For example, scenarios like the one studied in \citet{chamandy21} in which one planetary companion plunges into the stellar core and may deposit sufficient energy or angular momentum to help the other companion survive. These multi-body interactions could significantly affect the post-CE architecture and merit targeted simulations.
Second, our treatment of the envelope's internal energy has so far relied on simplified prescriptions. A more detailed approach, following methods such as those in \citet{Lagos21}, may improve estimates of the binding energy and shift the predicted survival boundary. 
Third, we note that our tidal evolution modeling has relied on two prescriptions and a single planet in the system. Additional work is needed to test the sensitivity of the planet's final orbit to different tidal formulations \citep{nord13,nord06}, including self-consistent dynamical-tide frameworks that couple stellar evolution to the full linear tidal response \citep{Narayan2026}, which can shorten engulfment timescales by factors of several relative to equilibrium tides.
Furthermore, additional planetary companions could be included in these simulations, potentially resulting in different system architectures. These alternatives could shift the range of parameter space in which outer planets remain bound and dynamically stable.

Lastly, our simulations have not considered accretion of the host star onto the planet and what that means for survival or detectability. Recent work by \citet{Jenkins26} simulated planet engulfment, concluding that the planet can accrete 3-48\% of its initial mass during this stage at an accretion efficiency of 1.0. This can be a significant increase which would allow initially lower-mass planets to become more massive and survive CE evolution. The cited work also predicts the observational signatures that can flag such accretion events. 

Future detections of wide-orbit planets around compact binaries via \textit{Gaia} astrometry, direct imaging, or transit timing could directly test these survival scenarios and calibrate key ingredients of common-envelope evolution and tidal models.

\section{Conclusion}\label{conclude}
In this work, we have simulated planetary systems via stellar evolution to understand the population of planets that survive common-envelope evolution. We explored a grid of simulation input parameters, varying assumptions such as host mass, tidal prescription, and envelope ejection efficiency. Our conclusions are as follows:

\begin{enumerate}
    \item Planets that survive CE evolution are rare. We tested 1\,$M_{\odot}$ and 2\,$M_{\odot}$ host star masses and static or variable tidal prescriptions for 10 million systems of each variation. We find that most planets fail to unbind the envelope before being destroyed.
    Our simulations for the most realistic values of $\alpha= 0.41$ or 1, based on population-wide estimates, result in a survival fraction of $\lesssim 10^{-7}$.
    In systems where we include equal envelope binding energy to gravitational energy ($\alpha = 2$), the survival fraction is $10^{-5}$. Finally, in the most optimistic case we considered ($\alpha=20$), the fraction of survivors is $10^{-3}$ to $10^{-2}$.
    \item CE survivors are preferentially massive, typically $\gtrsim 10\,M_{\rm{Jup}}$. High-mass planets are more effective at transferring energy and angular momentum during inspiral. Lower-mass planets rarely halt inspiral in time, suggesting that CE survival selects for the most massive end of the planet distribution.
    \item CE survivors end up in close-in orbits (within 1.4 tidal disruption radii) around the resulting white dwarf. This is in contrast to planets that undergo high-eccentricity migration, which typically end up in orbits $>2a_{{disrupt}}$ \citep{FordRasio06}. 
    The very close orbits of post-CE planets are therefore a distinct observational signature that could help differentiate them from planets that migrate via other mechanisms.

\end{enumerate}

The properties of the known planets around white dwarfs WD 1856b and WD 0310b stand at odds with our simulated CE outcomes. Their orbital separation and masses lie outside the parameter space of our typical survivors. While CE evolution \textit{can} explain the orbits of these planets (when additional energy sources are included), if this mechanism were operating we would expect to have also seen a larger and easier to detect population of massive planets with even smaller orbital separations. Since no such ultra-closely-orbiting massive planets around white dwarfs are yet known, our results suggest that the majority of close-in planets around white dwarfs arise from mechanisms other than CE evolution, such as high eccentricity migration.

\begin{acknowledgments} We thank Hailiang Chen and Smadar Naoz for helpful conversations. This research has made use of NASA's Astrophysics Data System Bibliographic Services. The authors acknowledge the MIT Office of Research Computing and Data for providing high performance computing resources that have contributed to the research results reported within this paper.

\software{\texttt{numpy} \citep{harris20}, \texttt{scipy} \citep{virt20}, \texttt{pandas} \citep{mckin10}, \texttt{matplotlib} \citep{hun07}, \texttt{MESA} \citep{pax11,pax13,pax15,pax18,pax19,jermyn23}.}
\end{acknowledgments}

\appendix \label{appendix:1}
\renewcommand{\theequation}{A.\arabic{equation}}
\setcounter{equation}{0}
We can derive an approximate expression to estimate whether or not a planet of mass $M_p$ will survive given a scenario with specific $\alpha, \lambda$ values, under the assumption that survival is determined primarily by whether the planet survives tidal disruption (as is typically the case for planets more massive than about 11 $M_{\rm{Jup}}$ in our simulations), rather than photoevaporation. 
Start with equation \ref{eqn:fin_af_sim} in the paper:
\begin{equation}
a_f = a_{\rm engulf} \frac{M_c}{M_\star} \left[\frac {2 M_{\rm env}}{\alpha \lambda r_L M_p} +1\right]^{-1}
\end{equation}
Because $M_{\rm env} \gg M_p$ we can neglect the ``1''.  This leads to
\begin{equation}
a_f \simeq a_{\rm engulf} \frac {\alpha \lambda r_L} {2 } \frac{M_c M_p}{M_\star M_{\rm env}} 
\label{eqn:afaiv2}
\end{equation}
Now, we make the assumption that the planet survives if it avoids tidal disruption. From equation (4) of \citet{Rappaport13} we can express the minimum separation before Roche-lobe overflow of a planet by 
\begin{equation}
a_f \gtrsim 2.34 \frac{(M_c/M_\odot)^{1/3}}{\rho_p^{1/3}} ~R_\odot
\end{equation}
where $M_c$ is the mass of the white-dwarf progenitor, and $\rho_p$ is the mean density of the planet in $g~cc^{-1}$.

If we substitute this result into equation (\ref{eqn:afaiv2}), and solve for $\alpha \lambda$ we find:
\begin{equation}
\alpha \lambda \gtrsim 4.67 \frac{M_g M_{\rm env}}{M_p M_c^{2/3}}\frac{1}{R_g}\frac{1}{\rho_p^{1/3}}
\end{equation}
where we now call the mass of the giant $M_g$ (same as $M_\star$, in $M_\odot$), we have replaced $a_{\rm engulf}$ with the radius of the giant, $R_g$ in $R_\odot$, when it starts to overfill its RL, and we have approximated $r_L$ as unity.  We can do this because the mass ratio ($M_g/M_p$) is so large that $R_L \simeq a_{\rm i}$.

The minimum value of $\alpha \lambda$ comes when $R_g$ is a maximum, which we can take to be $\sim$400 R$_\odot$ for a 1-3 M$_\odot$ star on the AGB.  If we take $M_g$ to be 1.5 M$_\odot$ as representative, and $M_c \simeq 0.6 M_\odot$ as a typical WD mass.  This comes to
\begin{equation}
\alpha \lambda \gtrsim 4.7 \, \frac{1.5 \times 0.8}{0.001(M_p/M_{\rm Jup}) \,0.6^{2/3} } \frac{1}{400} \frac{1}{1}
\end{equation}
where we have taken the mean density of gas giants to be near unity.

This leads to a simple analytic approximation for a planet to successfully eject the common envelop of its host star:
\begin{equation}
\alpha \lambda M_p \gtrsim 20 \, M_{\rm Jup}
\end{equation}
This result is broadly consistent with our findings using more detailed calculations, and provides an analytic estimate of the planetary masses we can expect to survive the common envelope.  Of course, the planet must also survive the heat bath of both the red giant interior and the nascent white dwarf if it is to last to the current epoch.

\bibliography{CE_latest2026}
\bibliographystyle{aasjournalv7}

\end{document}